\documentclass[reprint,nofootinbib,amsmath,amssymb,aps,floatfix]{revtex4-2}
\usepackage[T1]{fontenc}
\usepackage{amsmath}
\usepackage{graphicx}
\usepackage{dcolumn}
\usepackage{bm}

\usepackage{orcidlink}
\usepackage{adjustbox}
\usepackage{booktabs}
\usepackage{tabularx}
\usepackage{aas_macros}
\usepackage{url}
\usepackage{bbold}

\usepackage{hyperref}
\hypersetup{
    colorlinks,
    linkcolor={red!50!black},
    citecolor={blue!50!black},
    urlcolor={blue!80!black}
}

\newcommand\omegaem{\Omega_{\rm EM}}

\newcommand\hhyp{\mathcal{H}}
\newcommand\chyp{\mathcal{C}}
\newcommand\rhyp{\mathcal{R}}
\newcommand\shyp{\mathcal{S}}
\newcommand\nhyp{\mathcal{N}}
\newcommand\dgw{d_{\rm GW}}
\newcommand\dem{d_{\rm EM}}
\newcommand\demi{d_{{\rm EM},i}}
\newcommand\midbar{\,|\,}
\newcommand\gj{G_{j}}
\newcommand\rdlr{r_{\rm DLR}}

\newcommand\rhatdlr{\hat{r}_{\rm DLR}}
\newcommand\epsout{\epsilon_{\rm out}}
\newcommand\epscat{\epsilon_{\rm cat}}
\newcommand\zspec{z_{\rm spec}}
\newcommand\zphot{z_{\rm phot}}
\newcommand\sigmaspec{\sigma_{z, {\rm spec}}}
\newcommand\sigmaphot{\sigma_{z, {\rm phot}}}

\newcommand\iomega{\mathcal{I}_{\Omega}}
\newcommand\idl{\mathcal{I}_{D_{L}}}
\newcommand\iomegadl{\mathcal{I}_{{\Omega},D_{L}}}
\newcommand\dl{D_L}

\newcommand\kms{\mathrm{km\,s^{-1}}}
\newcommand\arcsec{^{\prime\prime}}
\newcommand{\farcs}{\mbox{$.\mkern-4mu^{\prime\prime}$}}

\begin{document}


\title{From Localization to Discovery: Bayesian Ranking of Electromagnetic Counterparts to Gravitational-Wave Events}

\author{Kendall Ackley\orcidlink{0000-0002-8648-0767}}
\email{kendall.ackley@warwick.ac.uk}
\affiliation{Department of Physics, University of Warwick, Gibbet Hill Road, Coventry CV4 7AL, UK}

\date{\today}

\begin{abstract}
The robust association of electromagnetic candidates discovered during follow-up of gravitational-wave alerts is challenging, not only due to the large sky areas and broad distance uncertainties, but also due to the tens to hundreds of unrelated optical transients that are observed per event. 
We present a Bayesian ranking framework to identify candidate electromagnetic counterparts to gravitational-wave events without requiring photometric or spectroscopic classification. The method ranks transient candidates according to their consistency with the three-dimensional gravitational-wave localization and quantifies the significance of their associations. The framework combines morphology-aware host association, physically motivated offset priors, host-inferred luminosity distance posteriors, peculiar velocity corrections, and a treatment for missing or incorrectly associated host galaxies. We apply the method to two events, GW170817 and GW190425. For GW170817, the method ranks AT2017gfo first and correctly recovers NGC\,4993 as the host galaxy. We find a candidate p-value of 0.030 for AT2017gfo and find none of the 32 background transient sources reach the ranking value of AT2017gfo. For GW190425, the highest-ranked candidate has a p-value of $5.88\times 10^{-4}$ relative to 5101 background sources. Correcting for the multiple comparisons problem using the full set of 433 candidates lowers the significance of the top-ranked candidate to a p-value of 0.255. The framework is directly applicable to transient candidates with only sky location information, enables rapid prioritization for more efficient follow-up, and leads to more reliable counterpart identification in current and future observing runs.

\end{abstract}


\maketitle

\section{Introduction} \label{sec:intro}

The LIGO-Virgo-KAGRA network of ground-based gravitational-wave (GW) interferometers has transformed astrophysics over the past decade. For multi-messenger astronomy, the joint detection of a binary neutron-star merger and its electromagnetic (EM) counterpart, GW170817/AT2017gfo, was decisive \citep{Abbott2017,Abbott2017GWGRB,Coulter2017}. The accompanying gamma-ray burst (GRB) and kilonova led to crucial results on $r$-process nucleosynthesis, constraints on the neutron-star equation of state, and an independent measurement of the Hubble constant \citep{Abbott2017Hubble,Abbott2017Kilonova,Abbott2017GWGRB,Abbott2017MMA,Abbott2019Properties,Abbott2018NSEOS}.

Despite the unprecedented impact of new physics following GW170817, the expansion of the second-generation GW detector network and a growing sample of neutron-star mergers reported in GWTC-4 \citep{GWTC4}, no other GW event with a definitive EM counterpart has been discovered. Several factors contribute to this. The properties of the merger can produce intrinsically faint EM emission, while viewing geometry can make the counterpart more difficult to detection. At the same time, increasing GW detector reach expands the observable volume while sky localizations remain broad. In practice, a single follow-up campaign can yield tens to hundreds of optical transients per alert where many, if not all, are unrelated to the merger.

Current follow-up strategies often prioritize the brightest or nearest galaxies within the GW localization volumes or restrict searches to the 90\% probability sky area \citep{Ackley2020, White2011, Hanna2014, Fan2014, Gehrels2016}. However, these choices do not fully exploit the three‐dimensional GW localization information \citep[e.g.,][]{Piotrzkowski2022, Veske2021}. As wide‐field optical surveys dedicated to GW follow-up, like the Gravitational-wave Optical Transient Observer \citep[GOTO;][]{Steeghs2022, Dyer2024}, improve in depth and cadence, and as the ground-based GW detector network continues to gain in sensitivity, the number of unrelated transients and candidate hosts will grow rapidly, further increasing the burden of false-positives.

Even with optimal scheduling, data collection, and processing, a major challenge remains. Many of the transients reported during follow-up are not promptly (or ever) spectroscopically classified. In fact, as of this publication date, only 5.68\% of TNS\footnote{\url{https://www.wis-tns.org/}} sources listed in association with O4 high-significance GW events have a reported classification \footnote{\url{https://www.wis-tns.org/ligo/events}}. Some transients are reported as single-epoch photometric data, while others may have sparsely sampled lightcurves. Therefore, the task is to identify which, if any, of the observed transients is the true EM counterpart to the GW event.

We address this with a Bayesian ranking framework that combines the three-dimensional GW skymap (sky position and luminosity distance) with the transient's sky position and an inferred EM distance obtained via host-galaxy association. The result is a statistic that provides a relative ranking of the joint association of each EM candidate to the GW event. We then quantify the significance of each joint association by comparing its ranking statistic against a background of unrelated sources constructed from transients reported before the GW event. For events with multiple candidates, we correct the p-value using the total number of candidates assessed.

Existing methods address parts of this problem. Spatial-only approaches associate transients (mainly supernovae) with host galaxies \citep{DELIGHT2022, Qin2022, ASTROPATH2021, BLAST2024arXiv, Gagliano2021, Bloom2002}. Host galaxy ranking schemes include redshift information but often treat luminosity distance as a single value \citep{Evans2016Follow, Evans2016Optim, EylesFerris2025, Bartos2015, Gehrels2016, Ducoin2020, Hanna2014, Singer2016, Arcavi2017, Antolini2017, Antolini2016b, Artale2020, Kanner2012, Salmon2020, Nuttall2010, Demasi2025}. Bayesian frameworks assess the coincidence significance of a single EM transient or combine distance posteriors with galaxy priors \citep{Fan2014, Ashton2018, Ashton2021, Piotrzkowski2022, MaganaHernandez2024, Veske2021, Mo2025}.

Complementary approaches for galaxy-targeted strategies have been developed for Swift XRT/UVOT follow-up of GW triggers \citep{Evans2016Follow, Evans2016Optim, EylesFerris2025}. These methods prioritize galaxies or observing fields by convolving the GW localization with galaxy catalog information, such as galaxy distance and luminosity, and catalog completeness. The framework presented here is complementary but operates at a later stage. That is, we begin from a set of transient candidates that have already been detected and marginalize over their possible host galaxies to rank the transient–GW associations.

To our knowledge, no existing approaches incorporate, within a single likelihood, physically motivated offset distributions with explicit use of host morphology, uncertainties in the peculiar velocity correction, uncertainties on incorrect photometric redshift assignment, and missing or incorrectly associated host galaxies from the catalog.

Here, we build upon these earlier approaches and develop a unified Bayesian ranking framework that treats these effects consistently. Given a GW skymap and a list of EM transients, the pipeline identifies candidate host galaxies, constructs and evaluates a transient-host likelihood that includes morphology-aware offsets and distance modeling in the local Universe (including frame conversion and peculiar velocity corrections), and marginalizes over the possible hosts to obtain a ranking statistic for each GW-EM association. We then quantify the significance of each association relative to the unrelated-source background described in Sec.~\ref{sec:pvalue}. We adopt a flat $\Lambda$CDM cosmology with $H_0=70\,\rm{km}\,{\rm s}^{-1}\,{\rm Mpc}^{-1}$ and $\Omega_{\rm m}=0.3$.

\section{Data}
\label{sec:implementation}

We use three inputs for the framework: (i) the public three-dimensional GW skymaps, (ii) a list of optical transient candidates that are spatially coincident with the GW alert, and (iii) galaxy catalogs that provide redshift and morphology information. For each catalog galaxy we standardize astrometry, convert measured redshifts to the CMB frame, and apply a peculiar velocity treatment to obtain a redshift posterior. We then transform these posteriors into $\dl$ distributions under a fiducial cosmology, which we use as the $\dl$ input to the distance overlap calculation.

\subsection{Gravitational Wave Skymaps}
\label{sec:data:gw}

We use the publicly available localization skymaps from GraceDB\footnote{\url{https://gracedb.ligo.org/}} that are generated by low-latency GW analysis pipelines such as \texttt{BAYESTAR} \citep{Singer2016Bayestar}, \textsc{Bilby}\citep{Ashton2019, RomeroShaw2020}, or \textsc{LALInference} \citep{Veitch2015}. The skymaps are provided in the standard \texttt{HEALPix FITS} format \citep{Gorski2005}. Each pixel contains the posterior probability density for sky position and the conditional distance distribution. In this work, we use the skymaps that are released in low-latency, or within the first few hours following the GW alert.

The skymaps can be complex in shape, often characterized as multi-modal, disjointed regions. For high signal-to-noise ratio events from a network of three detectors, they are typically more compact and may be well approximated as a single mode. To use the skymap in our likelihood, we represent the localization as a smooth three-dimensional density $p(\alpha,\delta,\dl\midbar\dgw)$ evaluated at each EM candidate's sky position and (inferred) luminosity distance. We compare several estimators for this density and detail their construction and relative performance in Sec.~\ref{sec:bayes}.

\subsection{Transient Sources}
\label{sec:tns_candidates}
We obtain transients from the collated lists of GW-associated transients on the Transient Name Server (TNS)\footnote{\url{https://www.wis-tns.org/ligo/events}}\footnote{\url{https://www.wis-tns.org/astronotes/astronote/2024-79}}. The service reports sources that lie within the 99\% GW localization region and includes objects discovered during the year preceding and during the two weeks following the GW trigger time ($t_{\rm GW}$). We refer to sources discovered before and after the GW trigger as the pre-merger sources and post-merger candidates, respectively. For each source, TNS provides the sky coordinates, discovery filter and magnitude, the time of first report, and any available classification. We restrict the sample to sources reported by TNS as optical transients and exclude non-optical events such as FRBs.

\subsection{Host Galaxy Catalogs}
\label{sec:catalogs}

We crossmatch each transient against a set of galaxy catalogs chosen for their sky coverage, completeness, and availability of redshift and morphology measurements. The catalogs are queried in a fixed priority order so that there is consistent behavior across each GW event.

\subsubsection{GLADE+}
GLADE+\footnote{\url{https://glade.elte.hu/}} \citep{Dalya2022} is a multi-survey compilation designed for GW follow-up. It combines data from GWGC \citep{White2011}, HyperLEDA \citep{Makarov2014}, 2MPZ \citep{Bilicki2014}, 2MASS-XSC \citep{Skrutskie2006}, WISExSuperCOSMOS \citep{Bilicki2016}, and SDSS-DR16Q \citep{Lyke2020}. It contains $\sim\!11.5$\,million galaxies and $\sim\!7.5\times 10^{5}$ quasars, and is approximately complete to $\dl\!\approx\!47$\,Mpc. It provides spectroscopic and photometric redshifts, as well as stellar mass and peculiar velocity estimates. 
Because GLADE+ does not supply uniform morphology over the catalog, we retrieve semi-major axis $a$, semi-minor axis $b$, and position angle $\phi$ (measured east of north) from the component catalogs of HyperLEDA and WISExSuperCOSMOS. For sources with matching WISExSuperCOSMOS identifiers, we obtain morphology from the SuperCOSMOS Science Archive (SSA)\footnote{\url{http://ssa.roe.ac.uk/index.html}} \citep{Hambly2001}.

\subsubsection{DESI Legacy Survey DR10}
The DESI Legacy Survey DR10 (LS-DR10)\footnote{\url{https://www.legacysurvey.org/dr10/}} \citep{Dey2019} provides deep $griz$ imaging over $\sim\!20{,}000\,\mathrm{deg}^2$ and contains $\sim\!3$\,billion unique sources, with photometric redshifts available for $\sim\!1.5$\,billion galaxies \citep{Li2024}. Galaxy morphologies are derived from \texttt{Tractor} model fits \citep{Lang2016ascl}, including half-light radii $r_{\rm eff}$, ellipticity components $(e_1,e_2)$, and position angles $\phi$ (measured east of north) from the best-fit models.

\subsubsection{Pan‐STARRS1 DR2}
Pan-STARRS1 DR2 (PS1-DR2)\footnote{\url{https://outerspace.stsci.edu/display/PANSTARRS/}} provides deep $grizy$ photometry and morphology for $\sim\!3\times 10^8$ sources over three quarters of the sky \citep{Chambers2016, Flewelling2020}. The catalog contains moments-based shape parameters, Kron radii, and model-based size estimates \citep{Magnier2020}. We use the second moments to derive the effective radius ($r_{\rm eff}$, the radius containing half the total flux), axis ratio ($b/a$, the ratio of semi-minor to semi-major axis), and position angle ($\phi$, measured east of north). 

\subsubsection{PS1-STRM and WISE-PS1-STRM}
The PS1-STRM value-added catalog\footnote{\url{https://archive.stsci.edu/hlsp/ps1-strm}} \citep{Beck2021} uses forced photometry from PS1-DR1 to classify $\sim\!3$\,billion sources and to provide photometric redshift estimates. The WISE-PS1-STRM extension\footnote{\url{https://archive.stsci.edu/hlsp/wise-ps1-strm}} \citep{Beck2022} crossmatches to WISE All-Sky \citep{Wright2010} in the infrared bands $W1$-$W4$. Although WISE-PS1-STRM contains only a subset of the sources in PS1-STRM ($\sim\!3.5\times 10^8$ entries), the galaxy-star-quasar classification and photometric redshifts are improved, particularly at high redshift and in dusty systems. Both catalogs provide shape or Kron‐like polarization parameters \citep{Kaiser1995} that we convert into semi‐axes and PA as above.

\subsubsection{NED} 
The NASA/IPAC Extragalactic Database (NED)\footnote{\url{https://ned.ipac.caltech.edu/}} is an aggregated repository of extragalactic objects from the literature, surveys, and catalogs. It offers extensive sky coverage and the most detailed morphology for well-studied galaxies. We query NED to update photometric or add missing redshifts from imaging catalogs using spectroscopic redshifts or velocity measurements ($cz$) converted to redshift where available. NED data is heterogeneous, and the morphology and redshift information are not standardized in the same way as LS-DR10 or PS1-DR2. Thus, we are cautious to update the photometric redshift results only for NED sources with spectroscopic redshift or velocity measurements or for when the photometric redshift results are missing or are deemed to be unreliable.

\subsubsection{NED‐LVS}
The NED Local Volume Sample (NED-LVS)\footnote{\url{https://ned.ipac.caltech.edu/NED::LVS/}} is a subset of NED which focuses on the nearby Universe ($\dl\!\lesssim\!1$\,Gpc) with direct relevance for GW follow-up \citep{Cook2023,NEDLVS}. As of June 2025, it contains $\sim\!2.1$\,million sources and reports near-infrared completeness of $\sim\!100\%$ at $\dl\!\sim\!30$\,Mpc and $\sim\!70\%$ at $\dl\!\sim\!300$\,Mpc. It provides redshifts or redshift-independent distances and includes stellar-mass and star-formation rate estimates.

\subsection{Catalog Queries}
For each GW event we query a fixed set of galaxy catalogs using a consistent set of criteria. The bulk of host galaxy candidates come from GLADE+, LS-DR10, and PS1-DR2. We then use NED, and NED-LVS where relevant, downstream to either update inaccurate redshifts or add in missing distance information. We show the full query recipes in Appendix~\ref{app:recipes}.

We impose two global constraints. First, we require a minimum line-of-sight luminosity distance $10$\,Mpc to exclude the very local volume where radial queries return impractically large numbers of low-probability sources and where catalog incompleteness in LS-DR10 and PS1-DR2 dominates. Second, we restrict the search to galaxies to a maximum projected physical separation of $R_{\perp, \max}=70\,\mathrm{kpc}$ from the transient position. This value is chosen based on observations and population synthesis modeling \citep{Fong2022, Wiggins2018}. For the initial cone, we convert $R_{\perp,\max}$ to an angular radius using the GW conditional distance distribution along the transient line-of-sight, as described in Sec.~\ref{sec:hostassoc} and retrieve all sources within that cone. For every galaxy returned by the query, we recompute the projected physical separation using its own distance.
Only galaxies with $R_{\perp}< 70\,\mathrm{kpc}$ receive non-zero host-association likelihood. If no such galaxy is found, the nearest galaxy is assigned for context, but it has zero host association likelihood.

\subsection{Redshift Frame and Peculiar Velocity Corrections}
\label{sec:data:redshifts}

To compare EM and GW distance information, we map galaxy redshift measurements to luminosity distance posteriors under a fiducial cosmology. Unless otherwise stated, catalog redshifts are treated as heliocentric. For GLADE+, we use the uncorrected heliocentric values. We transform all redshift measurements and their uncertainties to the CMB frame using the Planck dipole, adopting a Solar System barycenter speed $v_{\rm CMB}=369.82\,\kms$ toward $(l,b)=(264.021^\circ, 48.253^\circ)$ in Galactic coordinates \citep{Planck2020}. We marginalize the line-of-sight peculiar velocities with a Gaussian prior of width $\sigma_v=250\,\kms$ which we propagate into the redshift uncertainty. The resulting redshift posteriors are mapped to $\dl$ and used as the EM distance term in the association likelihood.

For each candidate host we construct a line-of-sight redshift distribution $p(z\midbar\gj, \dem)$ (see Sec.~\ref{sec:redshifts}). If a spectroscopic redshift (or velocity $cz$) is available, we apply the same CMB-frame transformation and peculiar velocity marginalization as described above. If only a photometric redshift is available, we propagate the reported uncertainty and model the distribution as a Gaussian mixture with both outlier and catastrophic outlier components (Sec.~\ref{sec:redshifts}). If no redshift is available, we adopt a prior uniform in comoving volume, $p(\dl)\propto \dl^2$, between $D_{L,\min}$ and a maximum distance horizon $D_{L,\max}$, which we transform to redshift and back to luminosity distance under the fiducial cosmology.

\section{Bayesian Ranking Framework}
\label{sec:bayes}

Our goal is to assign each transient candidate $T_i$ a single statistic that quantifies how strongly the combined GW and EM data favor a common-source interpretation $\mathcal{H}_\mathcal{C}$ over random chance coincidence $\mathcal{H}_\mathcal{R}$. We compare
\begin{itemize}
\itemsep0em 
\item $\mathcal{H}_\mathcal{C}$:\ $T_i$ is the true EM counterpart to the GW event
\item $\mathcal{H}_\mathcal{R}$:\ $T_i$ is an unrelated, random coincidence.
\end{itemize}
We refer to the GW dataset by $\dgw$, given as the three-dimensional skymap, and the EM dataset for candidate $T_i$ by $\demi$ given as the transient position and the information derived from host association. For brevity, we omit the index $i$ when the notation is unambiguous. 

We write the common source parameters as $\Theta=\{\Omega,\dl\}$ with sky position $\Omega$ and luminosity distance $\dl$. If multiple hosts $\gj$ exist for $T_i$, we marginalize $p(d_{\mathrm{EM},i}\midbar\Theta)$ over $j$ with weights given by the morphology-aware association likelihood (as described in Sec.~\ref{sec:hostassoc:like}).

Our approach builds upon the Bayesian framework of \cite{Piotrzkowski2022, Ashton2018} \citep[see also][]{Stachie2020}. The posterior odds ratio between $\mathcal{H}_\mathcal{C}$ and $\mathcal{H}_\mathcal{R}$ given the observed data is
\begin{equation}
\begin{aligned}
    \mathcal{O}_{\chyp/\rhyp} &= \dfrac{P(\hhyp_\chyp\midbar\dgw, \dem)}{P(\hhyp_\rhyp\midbar\dgw, \dem)} \\
    &= \dfrac{P(\dgw, \dem \midbar \hhyp_\chyp)}{P(\dgw, \dem \midbar \hhyp_\rhyp)}\times\dfrac{P(\hhyp_\chyp)}{P(\hhyp_\rhyp)}
\label{eq:postodds}
\end{aligned}
\end{equation}
where the first factor is the Bayes factor ($\mathcal{B}_{\chyp/\rhyp}$) and the second is the prior odds ratio ($\pi_{\chyp/\rhyp}$).

Under the common-source hypothesis, the GW and EM data share the common parameters $\theta$, and the joint likelihood is the marginalization over the parameters,
\begin{equation}
\begin{aligned}
    & P(\dgw,\dem \midbar \hhyp_\chyp) \\
    &=\int_{\Theta} P(\dgw\midbar\theta,\hhyp_\chyp)\,
                    P(\dem\midbar\theta,\hhyp_\chyp)\,
                    \pi(\theta\midbar\hhyp_\chyp)\, d\theta.
\label{eq:commonhypothesis}
\end{aligned}
\end{equation}
We define the alternative hypothesis $\hhyp_\rhyp$ as a composite hypothesis of signal and noise event hypotheses for each dataset \citep{Stachie2020, Piotrzkowski2022},
\begin{equation}
    P(\dgw, \dem \midbar \hhyp_\rhyp)
    = \!\!\!\sum_{X,Y \,\in\, \{\shyp,\nhyp\}}\!\!\! P(\dgw, \dem \midbar \hhyp_{XY}),
\label{eq:alt_hypothesis}
\end{equation}
where $X$ and $Y$ index whether each dataset is a true astrophysical signal ($\shyp$) or a noise event ($\nhyp$). We note that even if both signals are true astrophysical signals, it does not necessarily imply that they are related. 

In this work, we assume that both datasets are genuine astrophysical signals. The relevant alternative to the common-source hypothesis is the unrelated-source hypothesis, $\hhyp_{\shyp\shyp}$, in which the GW and EM events are real but arise from distinct astrophysical sources. For the ranking statistic used here, we have $\hhyp_\rhyp \approx\hhyp_{\shyp\shyp}$. Following \citet{Piotrzkowski2022} and \citet{Ashton2018}, the Bayes factor reduces to a \textit{posterior overlap integral} over the shared parameters $\theta$,
\begin{equation}
    \mathcal{B}_{\chyp/\shyp\shyp} \simeq \mathcal{I}_{\theta} = \int_{\Theta}
    \dfrac{p(\theta\midbar \dgw,\hhyp_{\shyp})\,p(\theta\midbar \dem,\hhyp_{\shyp})}{\pi(\theta)} \, d\theta ,
\label{eq:bcss}
\end{equation}
where $\pi(\theta)$ is the common prior over the shared source parameters. Alternative null hypotheses, in which either dataset is noise or otherwise non-astrophysical, can be included using Eq.~\eqref{eq:alt_hypothesis}, but they are not the focus of this work \citep[see][their Eqs.~13--17]{Ashton2018}.

The prior odds ratio $\pi_{\chyp/\shyp\shyp}$ depends on the joint GW-EM rates, the background rate of unrelated astrophysical transients, observational selection effects, survey cadence, and co-observing time windows \citep{Ashton2018, Ashton2021, Piotrzkowski2022}. Each of these inputs is uncertain and can introduce selection biases, particularly when comparing candidates found from separate surveys or follow-up strategies.

In this work, we do not attempt to calibrate these rate and selection terms. For a GW event and a list of EM candidates, the GW posterior is held fixed for all candidate evaluations, while EM information such as transient class, survey selection function, discovery efficiency, and background transient rate are not modeled explicitly. Therefore, we use $\mathcal{I}_{\theta}$ as a prioritization, or ranking, statistic rather than as a fully calibrated posterior odds of Eq.~\eqref{eq:postodds}.

The value $\mathcal{I}_{\theta}$ by itself does not quantify the statistical significance of an association. Even when all candidates are unrelated to the GW event, there will always be one source that will have the largest value of $\mathcal{I}_{\theta}$. We estimate an empirical p-value by comparing a candidate's ranking statistics with the distribution of values from sources unrelated to the GW event. We use the set of transients observed before the GW trigger as the set of unrelated sources, as these cannot be counterparts to the subsequent GW event. The construction of the background population, statistical interpretation, and treatment of multiple GW candidates are described in Sec.~\ref{sec:significance}.

With $\Theta = \{\Omega,\dl\}$ the sky-distance posterior overlap for each transient candidate is
\begin{equation}
\iomegadl = \int \dfrac{p(\Omega,\dl \midbar \dgw)\, p(\Omega,\dl \midbar \dem)} {\pi(\Omega,\dl)} \,d\Omega\,d\dl ,
\label{eq:jointoverlap}
\end{equation}
where $\pi(\Omega,\dl)=\pi(\Omega)\pi(\dl)$ is the separable joint prior. The GW and EM posteriors can be written in conditional form as,
\begin{equation}
\begin{aligned}
 p(\Omega,\dl \midbar \dgw) &= p(\Omega\midbar\dgw)\, p(\dl\midbar\Omega,\dgw) \\
 p(\Omega,\dl \midbar \dem) &= p(\Omega\midbar\dem)\, p(\dl\midbar\Omega,\dem)\, .
\label{eq:posterior_conditional}
\end{aligned}
\end{equation}
Under these assumptions, Eq.~\eqref{eq:jointoverlap} reduces to an approximation of factorized sky and distance terms
\begin{equation}
\begin{aligned}
\iomegadl &\simeq \dfrac{p(\hat\Omega\midbar\dgw)}{\pi(\hat\Omega)}
    \int \dfrac{ p(\dl\midbar\hat\Omega,\dgw)\, p(\dl\midbar\hat\Omega,\dem)}{\pi(\dl)}d\dl \\
    & \simeq \iomega\times\idl,
\label{eq:jointoverlap_delta}
\end{aligned}
\end{equation}

\subsection{Spatial Overlap Integral}
\label{sec:spatialoverlap}

The spatial overlap integral is
\begin{equation}
\iomega =\int \dfrac{p(\Omega\midbar\dgw)\, p(\Omega\midbar\dem)}{\pi(\Omega)} \; d\Omega
\label{eq:spatialoverlap}
\end{equation}

For an optical transient $T_i$ localized to sub-arcsecond precision, the astrometric uncertainty is negligible compared to the angular scale of the GW localization. Therefore, we approximate the EM posterior as a delta function at the observed transient position $\hat\Omega$
\begin{equation}
    p(\Omega\midbar\dem) \simeq \delta(\Omega-\hat\Omega),
\end{equation}
and evaluate the EM distance posterior only along this line-of-sight,
\begin{equation}
    p(\Omega,\dl\midbar\dem)
    \simeq
    \delta(\Omega-\hat\Omega)\,
    p(\dl\midbar\hat\Omega,\dem).
\end{equation}
Thus, Eq.~\eqref{eq:spatialoverlap} reduces to
\begin{equation}
\iomega \simeq \dfrac{p(\hat\Omega\midbar\dgw)}{\pi(\hat\Omega)}.
\label{eq:spatialoverlap_reduced}
\end{equation}
When the sky prior is uniform, $\pi(\Omega)=1/(4\pi)$, the denominator is constant across candidates within a single event and $\iomega \propto p(\hat\Omega \midbar \dgw)$.

If the reported astrometric uncertainty is not negligible, as may be expected from radio transients or sources in extremely crowded fields, we simply replace the Dirac delta function by a small kernel on the sphere, $p(\Omega\midbar \dem)=\mathcal{K}(\Omega;\hat\Omega,\Sigma_\Omega)$, so that $\iomega$ evaluates to the GW skymap convolved with $\mathcal{K}$ at $\hat\Omega$. For all of the optical candidates considered here, we use the Dirac delta approximation.

\subsection{Distance Overlap Integral}
\label{sec:distanceoverlap}

The distance overlap integral is
\begin{equation}
\idl = \int\dfrac{p(\dl\midbar\hat\Omega,\dgw)\, p(\dl\midbar\hat\Omega,\dem)}{\pi(\dl)} \;d\dl,
\label{eq:distoverlap}
\end{equation}
where $p(\dl\midbar\hat\Omega,\dgw)$ is the GW conditional distance posterior along the line-of-sight of the transient $T_i$, $p(\dl\midbar\hat\Omega,\dem)$ is the EM distance posterior, and $\pi(\dl)$ is the common prior on luminosity distance. In the absence of a known redshift to a transient, the EM distance posterior is inferred by associating the transient with a candidate host (Sec.~\ref{sec:hostassoc}). The EM distance posterior is then given as a host-marginalized distance posterior over all possible cataloged hosts.

\subsubsection{GW distance posterior}
We use the public three-dimensional GW skymap to obtain the joint posterior, $p(\Omega,\dl \midbar \dgw)$, and evaluate it along the line-of-sight to each transient. In practice we compute two quantities: the angular posterior density $p(\hat\Omega\midbar\dgw)$ evaluated at the transient position $\hat\Omega$ and the conditional line-of-sight distance posterior $p(\dl\midbar \hat\Omega, \dgw)$. We can evaluate the posterior using two equivalent strategies, kernel density estimation (KDE) or pixel-wise evaluation on the \texttt{HEALPix} grid.

For the KDE approach, we fit a kernel density to the posterior samples recovered from the GW skymaps $(\Omega,\dl)$ and evaluate the continuous density at $(\hat\Omega,\dl)$ as
\begin{equation}
    p_{\rm KDE}(\dl\midbar\hat\Omega,\dgw) =\frac{\sum_i w_i(\hat\Omega)\,\mathcal{K}_{\lambda}(\dl - D_{L,i})}{\sum_i w_i(\hat\Omega)},
\end{equation}
where $\mathcal{K}_{\lambda}$ is a Gaussian kernel of bandwidth $\lambda$ and $w_i(\hat\Omega)$ are the weights of each sample. We construct this density using the standard Gaussian KDE from \texttt{scipy} and a clustered kernel fit jointly in all three dimensions using \texttt{ClusteredKDE} as implemented in \texttt{ligo.skymap}\footnote{\url{https://git.ligo.org/lscsoft/ligo.skymap}}. The KDE redistributes probability into neighboring pixels, thus, the point-wise angular density $p(\hat\Omega \midbar \dgw)$ can be slightly lower than the posterior contained within a single \texttt{HEALPix} pixel, especially for skymaps with very narrow features.

Alternatively, we can evaluate the skymap at the \texttt{HEALPix} pixel containing $\hat\Omega$. This is the default in \texttt{RAVEN}\footnote{\url{https://git.ligo.org/lscsoft/raven}}, which is a pipeline that searches for coincidences between transient alerts and GW events. For a transient at $\hat\Omega$, we identify the corresponding skymap pixel $s$ and use the conditional distance ansatz stored in the public three-dimensional skymap,
\begin{equation}
\begin{aligned}
p(\dl\midbar\hat\Omega,\dgw) &= \mathcal{N}(\dl; \mu_s,\sigma_s,N_s)\,\dl^2 \\
&= \dfrac{\dl^2\,N_s}{\sqrt{2\pi}\sigma_s}\exp\left[- \dfrac{(\dl - \mu_s)^2}{2\sigma_s^2}\right],
\end{aligned}
\label{eq:gw_dl_pixel}
\end{equation}
for mean $\mu_s$, scale $\sigma_s$, and normalization $N_s$ at each pixel $s$ \citep[see][their Eq.~1]{Singer2016}. 
The corresponding joint probability that the source lies within pixel $s$ and at a 
distance between $\dl$ and $\dl + d\dl$ is 
\begin{equation}
    P(s,\dl \midbar \dgw) \,d\dl= \rho_s\,p(\dl\midbar\hat\Omega_s,\dgw)\,d\dl,
\end{equation}
where $\rho_s\equiv\rho(\hat\Omega)$ is the posterior probability contained in pixel $s$ \citep[Eq.~2,][]{Singer2016}. Here we use the conditional distance distribution in the distance-overlap term because the angular probability is accounted for separately through $\iomega$. Because there is no smoothing step in the evaluation, this approach will capture any sharp probability spikes in the localization.

For compact localizations such as GW170817, where the probability is concentrated over a small region, the KDE and pixel-wise evaluations return consistent results for $p(\hat\Omega)$ and $ p(\dl\midbar \hat\Omega)$. For broad or multi-modal localizations such as GW190425, the main limitation of the KDE approach is how it can effectively capture multi-modal angular structure, as have been noted in previous coincidence analyses \citep[e.g.][]{MaganaHernandez2024}. We use the clustered KDE implemented in \texttt{ligo.skymap} which is well-suited to the multi-modal skymaps as it identifies and evaluates distinct clusters of modes independently.

\subsubsection{EM distance posterior}
In the absence of an observed redshift for the transient, we query the set of galaxy catalogs to obtain a set of candidate hosts $\mathcal{G}_i=\{ \gj \}_{j=1}^N$, where $\gj$ is the $j$-th cataloged galaxy in the search region. We define $\varnothing$ for the case where the true host of $T_i$ is not present in $\mathcal{G}_i$.

The EM distance posterior is found by marginalizing over the possible host associations and missing host term,
\begin{equation}
\begin{aligned}
p(\dl \midbar \hat{\Omega},\dem) &= \sum_{j=1}^{N} p(\dl \midbar \gj,\dem)\,
p(\gj\midbar\dem) \\
&\;\quad \; +
p(\dl \midbar \varnothing,\dem)\,
p(\varnothing\midbar\dem),
\label{eq:em_distance_host_marginalization}
\end{aligned}
\end{equation}
where $p(\gj \midbar \dem)$ is the posterior probability that galaxy $\gj$ is the true host (Sec.~\ref{sec:hostassoc}) and $p(\varnothing\midbar \dem)$ is the posterior probability that the host is missing from the catalog (Sec.~\ref{sec:missinghosts}).
The host-association probabilities satisfy
\begin{equation}
\sum_{j=1}^{N} p(\gj\midbar \dem) + p(\varnothing\midbar \dem) =1.
\label{eq:host_nohost_normalization}
\end{equation}

Substituting Eq.~\eqref{eq:em_distance_host_marginalization} into Eq.~\eqref{eq:distoverlap} gives
\begin{equation}
\idl = \sum_{j=1}^{N} p(\gj\midbar\dem)\, \mathcal{I}_{\dl,j} + p(\varnothing\midbar\dem) \, \mathcal{I}_{\dl,\varnothing},
\label{eq:distoverlap_expanded}
\end{equation}
where the distance overlap for each galaxy in the search cone is
\begin{equation}
\mathcal{I}_{\dl,j} = \int \dfrac{ p(\dl\midbar\hat\Omega,\dgw)\, p(\dl\midbar \gj,\dem)}{\pi(\dl)} \,d\dl .
\label{eq:per_host_distance_overlap}
\end{equation}
In the case where there is no host galaxy identified, and thus, no informative host distance constraint, we take the EM distance posterior to be the same as the common distance prior $p(\dl\midbar\varnothing,\dem)=\pi(\dl)$. The distance prior then cancels in the overlap integral, giving
\begin{equation}
\mathcal{I}_{\dl,\varnothing} = \int p(\dl\midbar\hat\Omega,\dgw) \,d\dl = 1.
\label{eq:nohost_distance_neutral}
\end{equation}
Thus, the missing host contribution neither rewards nor penalizes any candidate based on whether a true host galaxy is found. Rather, this allows the candidate ranking to be set by the sky overlap $\iomega$ and by the posterior probability assigned to the missing host hypothesis.

The cancellation of the prior in Eq.~\eqref{eq:nohost_distance_neutral} is exact only for the deliberately uninformative case, i.e., where there are no additional distance-dependent host weights applied, such as weighting by luminosity, stellar mass or star formation rate weighting. If these are included, the missing host contribution must be modified in a similar manner. In particular, the catalog completeness term that enters $P_{\rm faint}$ should be weighted by the same host property, so that the missing host contribution represents the fraction of the weighted host population that is absent from the catalog.

The final host-marginalized distance overlap integral is
\begin{equation}
\idl = \sum_{j=1}^{N} p(\gj\midbar\dem)\, \mathcal{I}_{\dl,j} + p(\varnothing\midbar\dem),
\label{eq:fulldistoverlap}
\end{equation}
This quantity is evaluated over all candidate hosts in $\mathcal{G}_i$, not only over the highest probability host.

\subsubsection{Distance priors}
We note that both the GW and EM posteriors must be constructed under the \emph{same} prior $\pi(\dl)$, both in their functional form and over the same bounds. Otherwise, the inverse prior in Eq.~\eqref{eq:distoverlap} will not cancel properly. We adopt a fixed prior $\pi(\dl)$ for all candidates and all events, so that $\idl$ is on a common scale throughout the analysis.

In addition, we must be cautious about our choice of prior bounds. We choose a broad prior in distance rather than an event-specific narrow interval around the averaged all-sky distance. If we were to choose the narrow event-specific bounds, then $\idl$ is sensitive to arbitrary normalizations. Instead, we set global bounds on the distance prior with values broad enough to contain all plausible distances we may use in our search. Here, we set $[D_{L,\min}, D_{L,\max}] = [10\,\mathrm{Mpc}, 10^4\,\mathrm{Mpc}]$. If one compares two models for the \emph{same} candidate, then the overall prior normalization cancels in a ratio of $\idl$ values and the prior bounds can follow closely to the GW conditional distance posteriors \citep[see][]{Piotrzkowski2022}. 

We assume a prior that is uniform in comoving volume and isotropic on the sky, truncated to $\dl\in [D_{L,\min},D_{L,\max}]$. In terms of redshift this is
\begin{equation}
\pi(z)\ \propto \frac{dV_c}{dz\,d\Omega} = \frac{c\,D_M^2(z)}{H(z)} 
\label{eq:priordist}
\end{equation}
where $D_M(z)$ is the transverse comoving distance and
\begin{equation}
H(z) = H_0\sqrt{\Omega_{\rm m}(1+z)^3 + \Omega_{\rm k}(1+z)^2 + \Omega_\Lambda}
\end{equation}
is the Hubble parameter \citep{Hogg1999}. The prior on luminosity distance then follows by a change of variables,
\begin{equation}
\pi(\dl)\ \propto\ \pi(z)\,\left|\frac{dz}{d\dl}\right|
\ =\ \frac{c\,D_M^2(z)}{H(z)}\,
  \left|\frac{dz}{d\dl}\right|,
\end{equation}
where $\dl(z)=(1+z)\,D_M(z)$ and with $\pi(\dl)=0$ outside $[D_{L,\min},D_{L,\max}]$.

For completeness, $D_M$ relates to the line-of-sight comoving distance
$D_C(z)=c\int_0^z\!\,H^{-1}(z')\,dz'$ via
\begin{equation}
D_M(z)=
\begin{cases}
\frac{D_H}{\sqrt{\Omega_k}}\sinh\!\big(\sqrt{\Omega_k}\,D_C/D_H\big), & \Omega_k>0,\\[3pt]
D_C, & \Omega_k=0,\\[3pt]
\frac{D_H}{\sqrt{|\Omega_k|}}\sin\!\big(\sqrt{|\Omega_k|}\,D_C/D_H\big), & \Omega_k<0,
\end{cases}
\end{equation}
where $D_H\equiv c/H_0$ is the Hubble distance. In a flat cosmology ($\Omega_k=0$), $D_M=D_C$ and 
\begin{equation}
\frac{d\dl}{dz}=(1+z)\frac{c}{H(z)}+D_C(z)
\end{equation}.

\subsection{Temporal overlap integral}
\label{sec:timeoverlap}
A temporal overlap integral $\mathcal{I}_t$ can be introduced by analogy with $\iomega$ and $\idl$. The time overlap integral is defined as 
\begin{equation}
    \mathcal{I}_t = \int\!\dfrac{p(t_c \midbar \dgw)\,p(t_c \midbar \dem)} {\pi(t_c)}\; dt_c,
\label{eq:timeoverlap}
\end{equation}
where $t_c$ is the common physical merger time. The GW trigger time has a timing uncertainty that is negligible compared with optical survey cadences. Thus, the posterior can be approximated as
\begin{equation}
p(t_c\midbar d_{\rm GW})\simeq \delta(t_c-t_{\rm GW}),
\end{equation}
and the temporal overlap integral reduces to
\begin{equation}
    \mathcal{I}_t \simeq \frac{ p(t_{\rm GW}\midbar \dem)}{\pi(t_{\rm GW})}.
\label{eq:timeoverlap_approx}
\end{equation}

As a toy model, suppose the only EM timing information is a last non-detection at $t_{\rm non}$ and a first detection at $t_{\rm det}$. If the transient could have occurred uniformly in that interval, then
\begin{equation}
    p(t_c \midbar \dem) =
    \begin{cases}
        (t_{\rm det} - t_{\rm non})^{-1}, & t_{\rm non} < t_c < t_{\rm det},\\
        0, & \text{otherwise},
    \end{cases}
\end{equation}
The temporal term is nonzero only if $t_{\rm GW}$ falls between the last non-detection and first detection. Thus, the discovery time alone is not sufficient to determine whether an optical transient is temporally consistent with the GW event. A source discovered at $t_{\rm GW}+2\,{\rm d}$ may still be consistent with the merger, even if the field was not observed earlier, while a source discovered at $t_{\rm GW}+0.5\,{\rm d}$ is less informative if the prior observing window was broad. A physically motivated $\mathcal{I}_t$ would require modeling the expected rise and decline of the source together with the cadence, depth, visibility, and selection effects of the observations.

This toy model also illustrates the sensitivity of $\mathcal{I}_t$ to the EM time prior. We consider two otherwise identical candidates first detected $\Delta t_{\rm disc}=0.5\,{\rm days}$ and $2\,{\rm days}$ after the GW trigger. If the temporal prior is a log-normal distribution with characteristic time $\tau_{\rm peak}=1.5\,{\rm d}$ and width $\sigma_{\tau}=0.7$, the temporal factor mildly favors the earlier candidate, with $\mathcal{I}_t(0.5,{\rm d})/\mathcal{I}_t(2,{\rm d})\simeq1.27$. Changing only the characteristic time to $\tau_{\rm peak}=0.5,{\rm d}$ gives a ratio of $\simeq28.4$, while $\tau_{\rm peak}=3,{\rm d}$ gives $\simeq 0.18$, reversing the preference toward the later candidate. Thus, even in this simplified example, the temporal weighting is strongly model-dependent. We therefore do not include $\mathcal{I}_t$ in the ranking statistic used here. However, the framework is compatible with such an extension as $\mathcal{I}_t$ enters as a multiplicative factor on the sky and distance overlaps.

\subsection{Host Galaxy Redshifts}
\label{sec:redshifts}

We obtain the EM distance posterior $p(\dl\midbar\gj,\dem)$ from the host galaxy redshift returned by our catalog query. The observed redshift $(z_{\rm obs})$ combines the cosmological redshift $(z_{\rm cos})$, due to the expansion of the Universe, with the line-of-sight peculiar velocity $(v_{\rm pec})$, due to peculiar motions induced by large-scale matter distribution. For non-relativistic velocities, we have
\begin{equation}
1 + z_{\rm obs} \;=\; (1+z_{\rm cos})\left(1+\frac{v_{\rm pec}}{c}\right),
\end{equation}
to which we approximate as 
\begin{equation}
z_{\rm obs} \;\approx\; z_{\rm cos} + \frac{v_{\rm pec}}{c}\, .
\label{eq:z_relation}
\end{equation}
Correcting for peculiar motion is especially important at low redshift ($z \lesssim 0.1$), where it can be the dominant error \citep[see, e.g.,][for a detailed discussion on this effect in GW cosmology]{Cozzumbo2025}. We must also account for measurement uncertainties on the observed redshift and correct for peculiar velocity, as otherwise this may lead to additional biases in our result \citep{Mukherjee2021}.

To convert from redshift $z_{\rm cos}$ to luminosity distance $\dl$, we complete two main steps (i) marginalize over the unknown peculiar velocity and (ii) perform a change of variables under our fiducial cosmology.

We first construct a redshift posterior for the cosmological component, $p(z_{\rm cos}\midbar \gj,\dem)$, by marginalizing over the unknown peculiar velocity with a Gaussian prior $v_{\rm pec}\sim\mathcal{N}(0,\sigma_v^2)$. In the low-velocity limit of Eq.~\eqref{eq:z_relation}, this is then a convolution of the measurement model for $z_{\rm obs}$ with a Gaussian prior of width $\sigma_v/c$
\begin{equation}
\begin{aligned}
p(z_{\rm cos}\midbar\gj,\dem) \propto& \int\!p\left(z_{\rm obs}\!\mid\!z_{\rm cos}\!+\!\frac{v_{\rm pec}}{c}\right) \\
&\hspace{2em} \times \mathcal{N}\!\left(v_{\rm pec}; 0, \sigma_v^2\right)\;dv_{\rm pec}.
\label{eq:em_cosmo}
\end{aligned}
\end{equation}

Finally, we map $z_{\rm cos}$ to luminosity distance $\dl$ with our fiducial cosmology using
\begin{equation}
p(\dl\midbar\gj,\dem)
= p\Big(z^{\ast}_{\rm cos}(\dl)\Big)
  \left|\frac{dz}{d\dl}\right|_{z^{\ast}_{\rm cos}(\dl)},
\label{eq:emdistpost_clean}
\end{equation}
where $z^{\ast}_{\rm cos}(\dl) \equiv \dl^{-1}(\dl)$ is the unique solution of $\dl(z^{\ast}_{\rm cos})=\dl$.

The conditional posterior over the observed redshift in Eq.~\eqref{eq:em_cosmo} is derived from the host galaxy measurement model. While a Gaussian can be used as an adequate approximation, we provide three categorical forms depending on the identified host galaxy from our analysis. These are (i) spectroscopic, (ii) photometric, and (iii) \emph{no reported} redshift. The specific likelihoods for these cases are detailed in the following sections.

\subsubsection{Host galaxy with spectroscopic redshift}
\label{sec:redshifts:prior:spec}
For a host galaxy $ \gj$ with a spectroscopic redshift measurement $\zspec$ and uncertainty $\sigmaspec$, we model the measurement as a narrow Gaussian centered on the redshift
\begin{equation}
p(z_{\rm obs}\midbar z_{\rm cos},{\rm spec})
= \mathcal{N}\!\big(z_{\rm obs};\, z_{\rm cos},\, \sigmaspec^2\big)\;.
\end{equation}
Marginalizing over the unknown line-of-sight peculiar velocity with a Gaussian prior $v_{\rm pec}\sim\mathcal{N}(0,\sigma_v^2)$ (as we showed in Sec.~\ref{sec:redshifts}) can be given as
\begin{equation}
p(z_{\rm cos}\midbar\gj,\dem) = \mathcal{N}\left(z_{\rm obs}; z_{\rm cos}, \sigma_z^2 + (\sigma_v/c)^2\right),
\label{eq:em_cosmo_pec}
\end{equation}
where the peculiar velocity uncertainty adds in quadrature to the measurement error. Unless otherwise noted, we adopt $\sigma_v=250\,\kms$. For galaxy clusters, larger values may be more appropriate.

\subsubsection{Host galaxy with photometric redshift}
\label{sec:redshifts:prior:phot}
Photometric redshifts are subject to both measurement scatter as well as occasional systematic misestimate biases. To capture these effects, we model the redshift posterior distribution for galaxies with photometric redshifts $\zphot$ and uncertainties $\sigmaphot$ as a three-component Gaussian mixture rather than as a single Gaussian.

While most of the $\zphot$ estimates are reasonably accurate, we expect some fraction to be moderately biased (\emph{outliers}, $\epsout$) due to template-fitting algorithm mismatches and a smaller fraction to be completely incorrect (\emph{catastrophic outliers}, $\epscat$). The latter outliers can be expected when spectral features like emission lines are completely misidentified and taken for absorption features, where then the redshift estimate would have no relation to the true value \citep{Ilbert2006}. 

If we ignore the contribution of outliers, then any catastrophic outliers will contribute to overly sharp, but \emph{incorrect} $\idl$ results. At the same time, if $\sigmaphot$ is arbitrarily inflated with an additional uncertainty, then $\idl$ will have reduced discriminatory power. 

We write our redshift likelihood as a mixture model to capture the behavior of the \textit{core}, \textit{outlier}, and \textit{catastrophic outlier} components given the estimates of the photometric redshift biases in the catalog as
\begin{equation}
\begin{aligned}
    p(z_{\rm cos} \midbar \gj, \dem) & = (1-\epsout - \epscat)\,p_{\rm core}(z_{\rm cos}) \\ 
    & \hspace{1em} + \epsout\,p_{\rm out}(z_{\rm cos}) + \epscat \,p_{\rm cat}(z_{\rm cos}).
\end{aligned}
\label{eq:em_mixture}
\end{equation}
All components are normalized over common redshift bounds throughout the analysis.

Following \cite{Sanchez2014} and \cite{Duncan2022}, we calibrate the outlier fractions $\epsout$ and $\epscat$ for LS-DR10 using spectroscopic-photometric crossmatches from the same survey. We define residuals as $\Delta z\!=\!\zphot\!-\!\zspec$ and the quantities
\begin{itemize}
\item mean bias $\mu = \langle \Delta z \rangle$
\item core scatter $\sigma_{\rm core}\!\equiv\sigma_{68}\!=\!(\Delta z_{84}-\Delta z_{16})/2$. We use the 68th‐percentile half‐width $\sigma_{68}$ instead of the standard deviation ($\sigma_{\Delta z}$) as the latter can be inflated by rare outliers.
\item outlier fraction $\epsout: 2\sigma_{68}\!<\!|\Delta z-\mu|\!<\!3\sigma_{68}$ for moderate biases 
\item catastrophic outlier fraction $\epscat:|\Delta z-\mu|\!>3\sigma_{68}$ for catastrophic failures.
\end{itemize}
For LS-DR10, we find a mean bias $\mu\!=\!0.0062$, $\sigma_{68}\!=\!0.0796$ (or standard deviation $\sigma_{\Delta z}\!=\! 0.1988$), $\epsout\!=\!0.0426$ and $\epscat\!=\!0.023$. We use $\sigma_{68}$ as a floor on the reported photometric-redshift uncertainty, such that
\begin{equation}
\sigma_{\rm core} = \max(\sigma_{\rm phot},\sigma_{68}).
\end{equation}
For the other catalogs, as there is no similar uniform spectroscopic-photometric training set, we set conservative values of $\mu\!=\!0.0$, $\sigma_{68}\!=\!0.08$, $\epsout\!=\!0.05$ and $\epscat\!=\!0.03$.

\begin{figure}[t]
  \centering
  \includegraphics[width=0.5\textwidth]{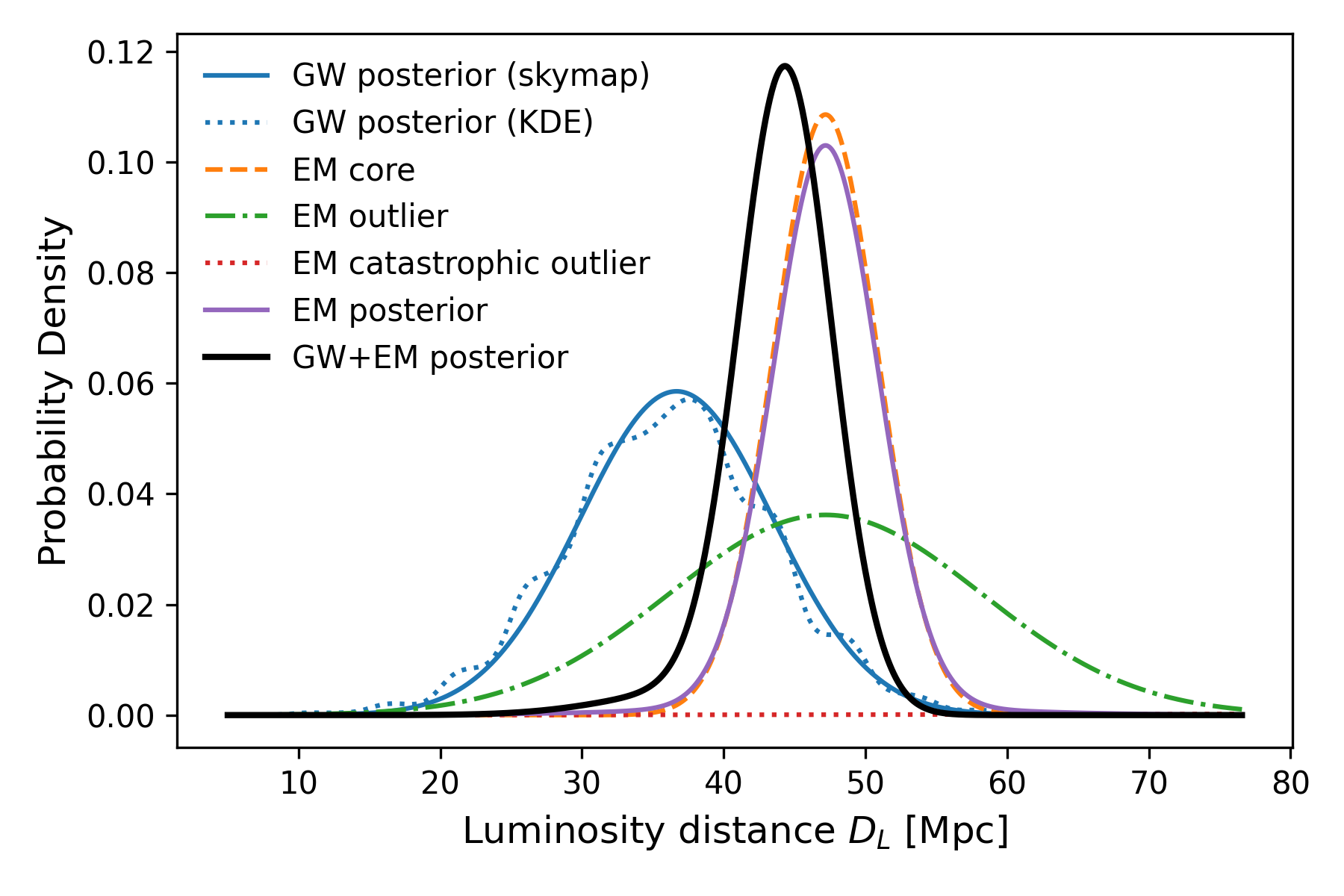}
  \caption{Illustration of the luminosity distance posteriors evaluated along the line-of-sight $\hat\Omega$ for AT2017gfo using the photometric redshift value from NED. The blue curve is the GW conditional luminosity distance from the public skymap (solid) and its KDE evaluation (dotted). The EM components show the host $\zphot$ mixture mapped to $\dl$ with peculiar velocity broadening for core (orange dashed), outlier (green dot dashed), and catastrophic outlier (red dotted). The purple solid curve is the resulting EM distance posterior for the full mixture. The black curve shows the normalized product proportional to the distance-overlap integrand of $\idl$ with the area equal to $\idl$.}
  \label{fig:dls_gw170817}
\end{figure}

The \textit{core} component captures the usual photometric scatter when the redshift estimation is successful and takes the form of a Gaussian centered on the bias-corrected photometric redshift,
\begin{equation}
    p_{\rm core}(z_{\rm cos}) = \mathcal{N}\bigl(z_{\rm cos}; (\zphot-\mu), \sigma_{\rm core}^2\bigr).
\end{equation}

The \textit{outlier} component accounts for the moderate biases that produce errors several times larger than the quoted uncertainty. We define this in terms of a broadened Gaussian distribution to capture the long tails
\begin{equation}
    p_{\rm out}(z_{\rm cos}) = \mathcal{N}\bigl(z_{\rm cos};(\zphot-\mu), (\xi\,\sigma_{\rm core})^2\bigr),
\end{equation}
where $\xi \gtrsim 3$ aims to capture the systematic errors that are $\sim3-5$ times larger than statistical uncertainties. We use $\xi\in[3,5]$.

The \textit{catastrophic outlier} component captures when the photometric solution is entirely incorrect and the reported $\zphot$ carries no reliable information about the true $z$. In this case, we choose a non-informative prior that the galaxy can have a redshift value anywhere within our horizon with equal probability per unit volume. 
Given that we assume galaxies are uniformly distributed in comoving volume, we have
\begin{equation}
p_{\rm cat}(z_{\rm cos}) =
    \frac{dV_c/dz}{\int_{z_{\min}}^{z_{\max}} (dV_c/dz')\,dz'} =
    \frac{dV_c/dz}{V_c(z_{\max})-V_c(z_{\min})}\,,
\label{eq:pcat}
\end{equation}
where $z_{\min}$ and $z_{\max}$ are the lower and upper bounds of the redshift interval. We set $z_{\min}=0$ and choose $z_{\max}$ from the smaller of the distance prior upper bound and the $3\sigma$ upper bound of the GW conditional distance posterior and $dV_c/dz$ is the differential comoving volume element \citep[][]{Hogg1999, RomeroShaw2020}.

We could alternatively truncate the distance prior based on the EM detection horizon. However, the GW distance posterior strongly suppresses contributions beyond $3\sigma$ and the choice of upper bound for the catastrophic-outlier prior has only a minor impact on the final ranking. Using the EM detection horizon would also introduce survey-specific dependencies, i.e. filter, depth, transient type, that we deliberately keep out of the location-only ranking statistic.

When marginalizing over the unknown line-of-sight velocity with $v_{\rm pec}\!\sim\!\mathcal{N}(0,\sigma_v^2)$, the Gaussian components variances will take on an additional component where
\begin{equation}
\begin{aligned}
\sigma_{z,{\rm core}}^2 &= (\sigma_{\rm core})^2 + (\sigma_v/c)^2 \\
\sigma_{z,{\rm out}}^2 &= (\xi\,\sigma_{\rm core})^2 + (\sigma_v/c)^2,
\end{aligned}
\end{equation}
while $p_{\rm cat}$ is unaffected.

In principle, the outlier fractions $\epsilon_{\rm out}$ and $\epsilon_{\rm cat}$ could be applied to each galaxy independently based on the number of photometric bands the galaxy has been detected in and the signal-to-noise ratio in each band, or based on statistical measures of the template fit. However, these parameters are not uniformly available across all catalogs and their calibration requires catalog-specific training sets. Therefore, we adopt the catalog-level outlier fractions for practicality and leave the evaluation of individual galaxy outlier probabilities for future work.

\subsubsection{Host galaxy with no reported redshift}
\label{sec:redshifts:prior:uninform}
For candidate hosts without a redshift measurement, we adopt an uninformative redshift prior with fixed bounds. We consider two standard choices: (i) uniform in comoving volume on $[0,z_{\max}]$ given in Eq.~\eqref{eq:pcat} or (ii) uniform in Euclidean volume
\begin{equation}
    p(z \midbar \dem) \propto z^2,
    \quad 0 < z < z_{\max}\;.
\label{eq:uniform_z2}
\end{equation}
As our analysis focuses on $z \lesssim 0.1$, Eqs.~\eqref{eq:pcat} and \eqref{eq:uniform_z2} differ at the $\sim$10\% level after mapping to $\dl$. Thus, either serves as a suitable uninformative prior when no galaxy redshift is available. We use the uniform in comoving volume prior for consistency with the global distance prior. We then convert to a $\dl$ prior via the usual change of variables under the fiducial cosmology.

Although the distance term is intentionally uninformative, the candidate host-transient pair can still be promoted (or downweighted) through other factors. The inclusion of this term can be interpreted as a way to keep the statistics consistent when applying additional astrophysical weights, such as stellar mass, star-formation rate, or luminosity. This is particularly useful when two candidates are spatially comparable but one lacks a redshift or has a photometric redshift so broad that it is effectively uninformative. This allows for the morphology or offset likelihood and any optional astrophysical weights to determine the ranking rather than a poorly constrained distance term.

\section{Transient-Host Galaxy Association}
\label{sec:hostassoc}

For each transient, we want to identify the most probable host among the galaxies within an angular search radius $\Delta\theta_{\max}$. For each catalog entry in the set of host candidates $\{\gj\}_{j=1}^N$, we compute the posterior probability that galaxy $\gj$ is the true host for $T_i$,
\begin{equation}
p(\gj\midbar \dem) = \frac{p(\dem \midbar \gj)\,\pi( \gj)}
{\sum_{k=1}^N p(\dem \midbar G_k)\,\pi(G_{k}) \,+\,p(\dem\midbar\varnothing)\,\pi(\varnothing)}\,,
\label{eq:hostmatch}
\end{equation}
where $p(\dem\midbar \gj)$ is the likelihood built from catalog information on the host centroid $\Omega_{j}$, redshift $z_j$, morphology parameters, and their associated uncertainties, $\pi(\gj)$ is the prior over the host galaxy candidates, and $p(\dem\midbar\varnothing)\pi(\varnothing)$ accounts for missing hosts in the catalog (Sec.~\ref{sec:missinghosts}). 
To identify the most probable host, we evaluate
\begin{equation}
    p_{\rm assoc} \equiv \max_{j}\,p(\gj\midbar \dem),
\label{eq:passoc}
\end{equation}
which is defined as the host association probability.

\subsection{Host Association Prior}
\label{sec:hostassoc:prior}
There is extensive literature on observable properties of transients as a function of their offsets from their hosts. These include systematic studies for GRBs \citep{Bloom2002, Fong2013, Blanchard2016, Fong2022, Mandhai2022, Nugent2022, OConnor2022, Zevin2022, Jeong2024, Gaspari2025}, galactic binary neutron stars \citep{ ODoherty2023, Disberg2024, Gaspari2024gal}, supernovae \cite{Hsu2024}, fast blue optical transients \citep{Chrimes2024}, and fast radio bursts \citep{Bhandari2022, Gordon2025, ASTROPATH2021}.

We adopt a uniform prior over the $N$ galaxies within the angular search region, 
\begin{equation}
   \pi(\gj)=\frac{1-\pi(\varnothing)}{N}. 
\label{eq:uniform_hostassoc_prior}
\end{equation}
This choice avoids coupling the host association to a specific population model and deliberately keeps it separate from an astrophysical event-rate weighting.

More informative priors can be used as physically motivated alternatives to the uniform prior of Eq.~\eqref{eq:uniform_hostassoc_prior}. A standard approach for selecting host galaxies to GW events is to assume that more luminous or more massive galaxies are intrinsically more likely to host the transient population, which also carries an intrinsic astrophysical rate assumption. This assumption can be applied to the prior as weights to the uniform prior with the weights defined depending on a specific choice of progenitor tracer, i.e., weighting by galaxy luminosity, metallicity, stellar mass, or star-formation rate, where near-infrared luminosity can proxy for stellar mass while optical luminosity more closely traces recent star-formation rate \citep[e.g.,][]{Evans2016Follow, Evans2016Optim, Ducoin2020, Palfi2025, Artale2020, EylesFerris2025}. 

A weighted prior can be introduced as
\begin{equation}
\pi(\gj)=(1-\pi(\varnothing)) \, \frac{W_j}{\sum_{k=1}^{N} W_k},
\label{eq:weighted_prior}
\end{equation}
where $W_j$ can be chosen to represent luminosity $W_j = L_j$, stellar mass $W_j = M^{\star}_{j}$, or star formation rate $W_j = \mathrm{SFR}_j$. These weightings are well-motivated astrophysically for a specific source population. However, the observable is inherently population-dependent. For example, weighting galaxies by stellar mass does not necessarily produce the same ranking as those weighted by star-formation rate.
In any case, the missing host and catalog completeness contributions in $\pi(\varnothing)$ of Eq.~\eqref{eq:weighted_prior} must have consistent weighting treatments applied. The effect of this is expected to be largest when several galaxies have similar offset and redshift support, but substantially different luminosities.

For our method, we do not assume any particular class of the transients when evaluating the ranking statistic. Therefore, we use equal prior weights for candidate hosts and the host association is determined by the morphology-aware offset likelihood and redshift information, if available.

\subsection{Host Association Likelihood}
\label{sec:hostassoc:like}
We model the galaxy likelihood in the numerator of Eq.~\eqref{eq:hostmatch} as the product of a spatial (astrometry and morphology) term and a redshift term,
\begin{equation}
\begin{aligned}
p(\dem\midbar \gj)
&= p(\rdlr \midbar \gj)\times p(z\midbar\gj)\\ 
& \equiv p_{\rm offset}(\gj)\,p_{z}(\gj).
\label{eq:likeli_galaxy}
\end{aligned}
\end{equation}
where $p_{\rm offset}(\gj)$ is the likelihood of observing the host-normalized offset between the transient and candidate host, and $p_z(\gj)$ is the likelihood of observing the transient redshift given the host redshift posterior. For transients without an independently measured redshift, we set $p_z(\gj)=1$ for the host-association step as the host redshift information is used in the distance overlap term of Eq.~\eqref{eq:distoverlap}. We treat these terms as separable because the astrometric and redshift uncertainties are modeled as independent measurement processes.

The offset parameter $\rdlr$ is a dimensionless, host-normalized offset,
\begin{equation}
r_{\rm DLR} \equiv \frac{\Delta\theta}{\mathrm{DLR}}.
\label{eq:rdlr}
\end{equation}
where $\Delta\theta$ is the projected angular separation between the transient and the galaxy centroid in the plane of the sky and DLR (\textit{directional light radius}) is the radius of a galaxy's elliptical light profile in the direction of the transient \citep{Gupta2016, Sullivan2006}. For an elliptical light profile with semi-major axis $a$, semi-minor axis $b$, and position angle $\phi$ (measured east of north), 
\begin{equation}
   \mathrm{DLR} = \dfrac{ab}{\sqrt{(a\sin\beta)^2 + (b\cos\beta)^2}}.
\label{eq:dlr}
\end{equation}
where $\beta$ is the angle between the galaxy's major axis and the vector connecting the host center to the transient. This morphology-based normalization allows us to compare offsets between hosts with varying sizes and ellipticities. These terms are computed from the galaxy morphology parameters provided in the respective catalogs.

\subsection{Angular Search Radius}
For each transient at sky position $\omegaem =\hat\Omega$ we search for candidate hosts within an angular radius determined from the maximum projected physical offset $R_{\perp,\max}=70\,{\rm kpc}$.
The physical offset scale is chosen to capture the majority ($\gtrsim90\%$) of expected compact-binary offsets based on population synthesis models and observed GRB host offset distributions \citep{Wiggins2018, Fong2022, Gaspari2025}. We define the lower line-of-sight distance used to set the query radius as
\begin{equation}
    D_{L,\rm \min}(\hat\Omega) = \max\!\left[\mu_{D_L}(\hat\Omega)-3\sigma_{D_L}(\hat\Omega),\, 10\,{\rm Mpc} \right],
\end{equation}
where $\mu_{D_L}(\hat\Omega)$ and $\sigma_{D_L}(\hat\Omega)$ are the parameters of the GW conditional distance distribution along the line-of-sight. The physical offset is converted to an angular radius
\begin{equation}
    \Delta\theta_{\rm phys} = \frac{R_{\perp,\max}}{D_A(z_{\min})},
\end{equation}
where the ratio is converted from radians to arcseconds, $D_A(z)$ is the angular-diameter distance \citep{Hogg1999} and $z_{\min}$ is converted from the $3\sigma$ lower bound of the GW conditional distance posterior in units of Mpc along the line-of-sight \citep{Singer2016}. We then find the angular radius in units of arcseconds
\begin{equation}
    \Delta\theta_{\max} \;=\; \max\left[\Delta\theta_{\rm phys}, 60\arcsec\right].
\end{equation}
We round $\Delta\theta_{\max}$ up to the nearest $10\arcsec$ to account for larger centroid errors and we impose a floor of $D_{L,\min}=10$\,Mpc to avoid excessively large cones in the very local volume.

For transients without a reported spectroscopic redshift, the redshift contribution is obtained by marginalizing over the GW distance/redshift posterior and the catalog redshift uncertainty. Here, the redshift term in Eq.~\eqref{eq:likeli_galaxy} reduces to the integrated prior, or $p(z\midbar\dem)=1$ leaving only the spatial offset term 
\begin{equation}
p(\dem\midbar \gj)
\approx p(\rdlr \midbar \gj).
\label{eq:spatial_off_gal}
\end{equation}

We define $r_0$ to be the \emph{true} but unknown host‐normalized offset and $\rhatdlr$ to be the observed host-normalized offset. The spatial (offset) likelihood is obtained by marginalizing over $r_0$,
\begin{equation}
p(\rhatdlr\midbar \gj)
=\sum_{c}\pi(c)\int_0^\infty\!p(\rhatdlr\midbar r_0,\sigma_{r,j})\, \pi(r_0\midbar c)\,dr_0,
\label{eq:hostoffset}
\end{equation}
where $\sigma_{r,j}\equiv \sigma_{\theta,j}/\mathrm{DLR}_j$ is the total astrometric uncertainty expressed in DLR units. 

\subsubsection{Astrometric Uncertainty Kernel}
\label{sec:hostassoc:astro}
Assuming that the positional uncertainties for the transient and galaxy centroids are independent, normally distributed, and positive, the observed radial separation in DLR units follows a Rice distribution \citep{Rice1945}
\begin{equation}
p(\rhatdlr\midbar r_0,\sigma_r)
=\frac{\rhatdlr}{\sigma_r^2}
\exp\!\left[-\frac{\rhatdlr^2+r_0^2}{2\sigma_r^2}\right]
I_0\!\left(\frac{\rhatdlr\,r_0}{\sigma_r^2}\right),
\label{eq:rice}
\end{equation}
where $I_0$ is the modified Bessel function of the first kind with order zero. If the true offset is exactly zero $r_0 = 0$, then this reduces to the Rayleigh distribution. 

\subsubsection{Intrinsic Offset Prior (Gamma Kicks)}
\label{sec:hostassoc:kick}
Compact-object binaries can be observed at large projected offsets from their formation sites because of natal kicks imparted during binary formation and evolution \citep{Gaspari2024, Gaspari2025, VignaGomez2018, Zevin2020, Zevin2022, Fong2022, Blanchard2016}. These kicks impart non-negligible and often modest velocities to the system. However, some systems can be kicked far from their hosts where they may be observed with asymmetric and long-tailed offset distributions. 

To capture a distribution with predominantly small displacements and a non-negligible tail for large offsets, we place a prior on the host-normalized intrinsic offset $r_0 > 0$ using a Gamma distribution
\begin{equation}
\pi(r_0\midbar c) =\frac{r_0^{\,k_c-1}e^{-r_0/\lambda_c}}{\Gamma(k_c)\,\lambda_{c}^{\,k_c}},
\label{eq:gamma_offset_prior}
\end{equation}
where $c$ indexes the assumed transient class or source population, $k_c$ is the shape parameter, and $\lambda_c$ is the scale parameter. For example, $c$ could distinguish populations with different intrinsic offset distributions, such as long or short GRBs, which typically show smaller and larger offset distributions, respectively \citep{Fong2022, Blanchard2016}. Since $r_0$ is normalized by the DLR, both $r_0$ and $\lambda_c$ are dimensionless and $\lambda_c$ does not explicitly depend on host size. Host morphology enters through the definition of $\rdlr$, rather than through an additional dimensional scale factor.

The shape parameter controls the concentration of the prior near the host center. For $k_c=1$, the Gamma distribution reduces to an exponential distribution with maximum density at $r_0=0$ and no preferred displacement. For $k_c>1$, the prior has zero density at the host center, peaks at a finite offset from the host center, and then decays exponentially. In this work, we adopt a single offset component with $k_c=2$ and $\lambda_c=1$, motivated by observed compact-binary and GRB host-normalized offset distributions \citep[e.g.][]{Bloom2002, Berger2010, Fong2013, Blanchard2016, Fong2022, OConnor2022}. This gives
\begin{equation} 
\pi(r_0\midbar G_j) \propto r_0 e^{-r_0}, 
\end{equation} 
with a peak at $r_0=1$ and an exponential tail to larger host-normalized offsets. The class index $c$ allows the offset prior to be replaced by a population-dependent prior when the transient class is known. An application of this class-dependent approach to the host association of GRB\,250818B is explored in \citet{Belkin2026}.

\begin{figure*}[!ht]
\includegraphics[width=0.3325\textwidth]{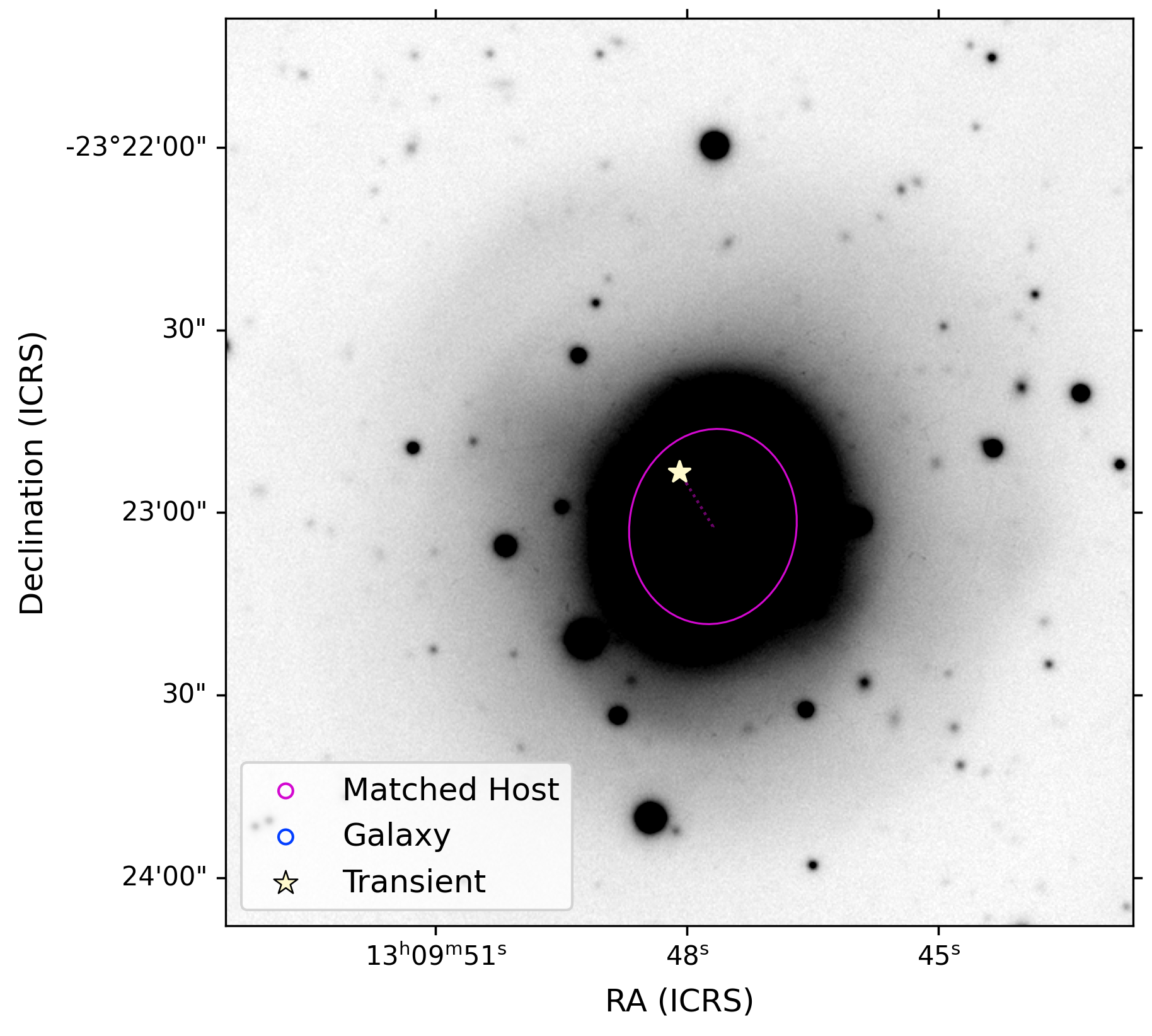} 
\includegraphics[width=0.32\textwidth]{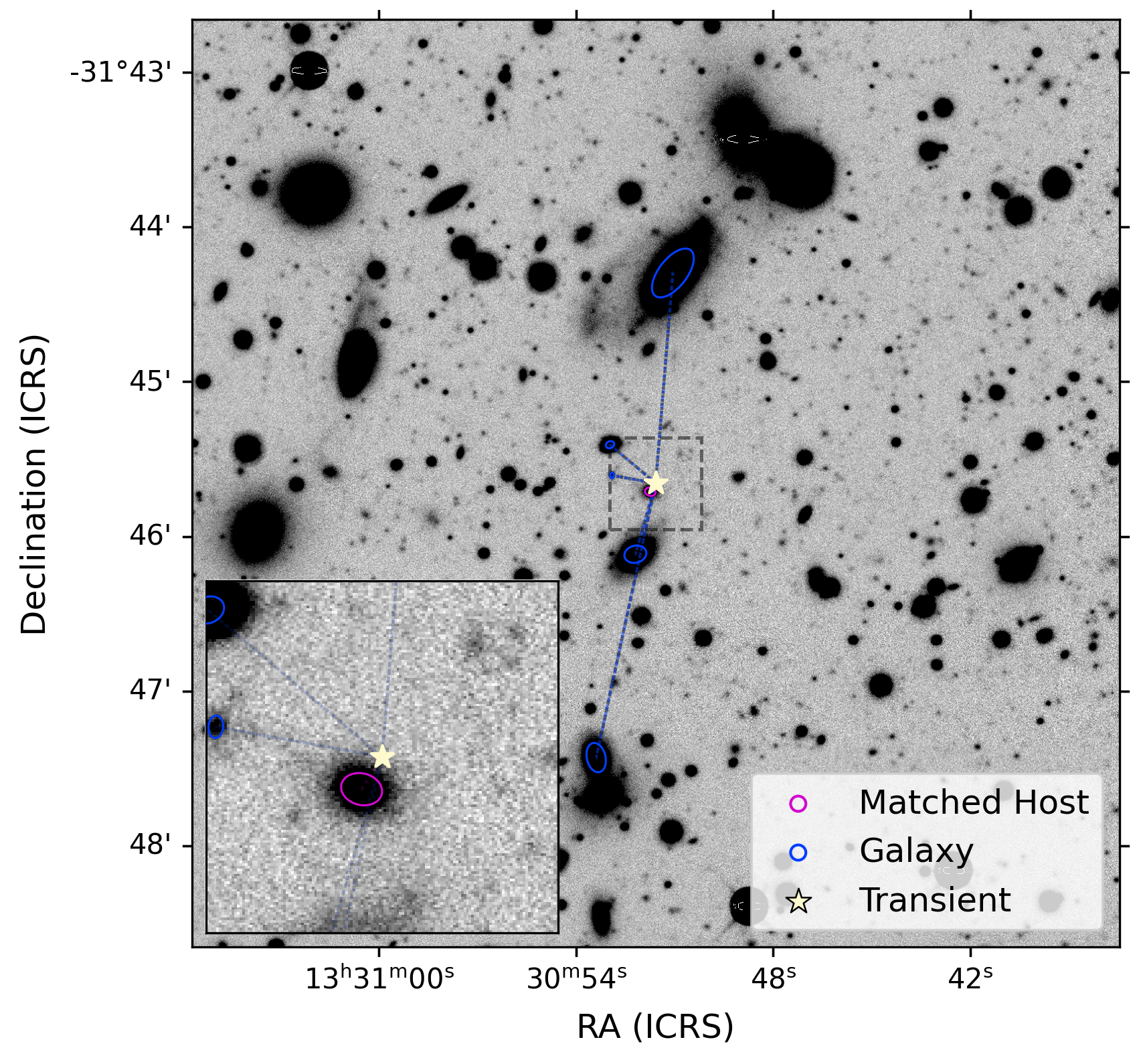}
\includegraphics[width=0.32\textwidth]{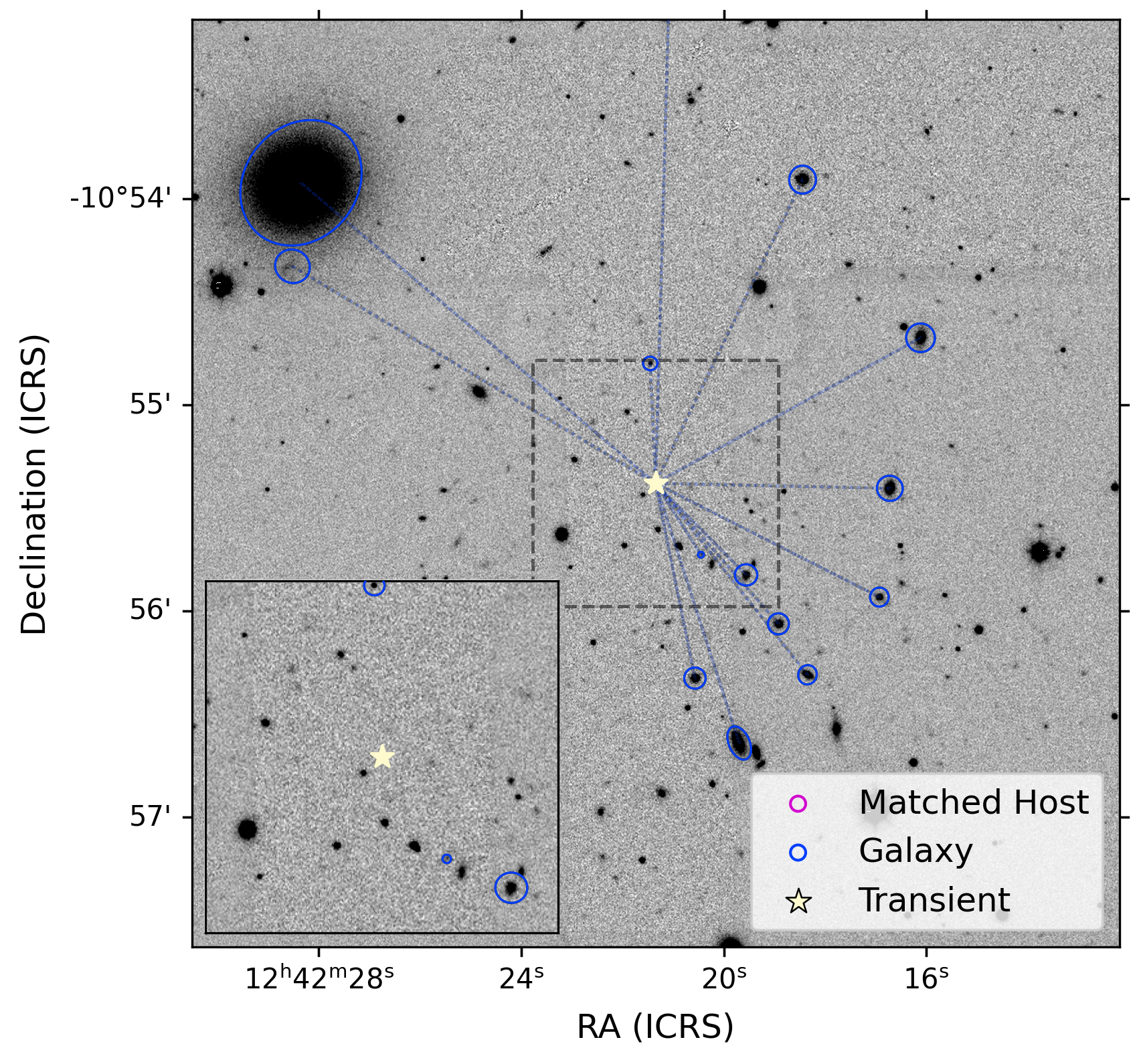}
\caption{Host association results from LS-DR10 for three GW170817 transients. The galaxies are those which lie within a $\rdlr \leq 30$ from the transient. The yellow star marks the transient and the magenta ellipse indicates the most probable host, if found. Blue ellipses mark all catalog galaxies within the projected physical offset. The shape parameters of the ellipse use the morphology parameters given in the catalog. The dotted vector points from the host centroid to the transient.
Left: AT2017gfo. The magenta ellipse shows the identification of the true host (NGC\,4993). 
Middle: AT2017bej. The lower-left zoom inset shows a clearer image of the geometry and offset of the preferred association. 
Right: AT2017arb. There is no host associated with this transient. The large galaxy to the upper left of the transient position has a projected physical offset of $R_\perp\simeq132\,\mathrm{kpc}$, outside of the cutoff $R_{\perp,\max}\leq 70\,\mathrm{kpc}$. The angular size of the $R_{\perp,\max}$ radius is $371\arcsec$ for AT2017gfo, $29\arcsec$ for AT2017bej, and $72\arcsec$ for AT2017arb.}
  \label{fig:hostassoc_dlr}
\end{figure*}

\subsection{Missing hosts in catalog}
\label{sec:missinghosts}

The denominator of Eq.~\eqref{eq:hostmatch} includes a ``no cataloged host'' term that accounts for the possibility that the \emph{true} host galaxy is not present in the queried catalog set. The posterior probability of the missing host hypothesis is given as the null probability, $p_{\varnothing} \equiv p(\varnothing\midbar\dem)$,
\begin{equation}
p(\varnothing\midbar\dem) = \frac{p(\dem\midbar\varnothing)\pi(\varnothing)}{
\sum_{k=1}^{N_i} p(\dem\midbar G_k)\pi(G_k) + p(\dem\midbar\varnothing)\pi(\varnothing)}\,,
\label{eq:p_empty}
\end{equation}
where $\pi(\varnothing)$ is the prior and $p(\dem\midbar\varnothing)$ is the likelihood of the observed EM data under the missing host hypothesis. We approximate the likelihood using first-order contributions from the main ways in which the true host can be absent from the cataloged host set,
\begin{equation}
p(\demi\midbar\varnothing) \simeq P_{\rm faint} + P_{\rm out} + P_{\rm rand} + P_{\rm mis}
\label{eq:null_weight}
\end{equation}
where $P_{\rm faint}$ is the probability that the true host is fainter than the catalog limit, $P_{\rm out}$ is the probability that the host lies outside the maximum search radius, $P_{\rm rand}$ is the probability that the apparent association is a chance foreground or background alignment, and $P_{\rm mis}$ accounts for catalog mis-classifications such as stellar contamination, spurious catalog entries, or deblending issues where fragments of a genuine host are treated as independent sources (see Sec.~\ref{sec:catalog_caveats}). We neglect the products of these terms as higher-order corrections, such as the probability that the host is both outside of the search radius and fainter than the catalog limit, as they are expected to be negligible.

\subsubsection{Faint or uncataloged hosts}

We set $P_{\rm faint}$ using the effective completeness of the host catalog along the transient line-of-sight. For a transient at sky position $\hat{\Omega}$, we write
\begin{equation}
\begin{aligned}
P_{\rm faint}(\hat{\Omega}) & =
\int \left[1-C_{\rm eff}(z,\hat{\Omega})\right]\,
p(z\midbar\hat{\Omega},\dgw)\,dz \\
& =\; 1 - \int C_{\rm eff}(z,\hat{\Omega}) \; p(z\midbar\hat{\Omega},\dgw)\,dz\,,
\label{eq:Pfaint_marginal}
\end{aligned}
\end{equation}
where $C_{\rm eff}(z,\hat{\Omega})$ is the effective catalog completeness, and $p(z\midbar\hat{\Omega},\dgw)$ is the GW conditional distance posterior expressed as a redshift probability density. The GW redshift density is determined through
\begin{equation}
p(z\midbar\hat{\Omega},\dgw) \propto p\!\left(\dl(z)\midbar\hat{\Omega},\dgw\right) \left|\frac{d\dl}{dz}\right|,
\label{eq:pgwz}
\end{equation}
with normalization
\begin{equation}
\int p(z\midbar\hat\Omega,\dgw)\,dz=1.
\end{equation}

The evaluation of $C_{\rm eff}$ depends on the catalog that supplies the host information. For magnitude-limited imaging catalogs such as LS-DR10 and PS1-DR2, we estimate the completeness from the limiting magnitude at the transient position. For LS-DR10, $m_{\rm lim}(\hat{\Omega})$ is computed from the inverse-variance depth map. For PS1-DR2, we use the survey-average stacked limiting magnitude in the selected band. For GLADE+, which is a heterogeneous galaxy compilation rather than a single imaging survey, we use the published catalog completeness curve \citep{Dalya2022}.

For magnitude-limited catalogs, the limiting apparent magnitude defines the faintest absolute magnitude detectable at redshift $z$,
\begin{equation}
M_{\rm thresh}(z,\hat{\Omega}) = m_{\rm lim}(\hat{\Omega}) - {\rm DM}(z) - K(z),
\end{equation}
where ${\rm DM}(z)$ is the distance modulus and $K(z)$ is the $K$-correction. We assume the reported magnitudes are already corrected for Galactic extinction. When robust $K$-corrections are unavailable, we estimate the systematic uncertainty by evaluating the completeness over the grid $K(z)\in\{-1,0,+1\}$ mag.

Following the general approach used for catalog-incompleteness corrections in dark-siren cosmology \citep[e.g.,][]{Turski2026,Borghi2026}, we model the host-galaxy luminosity distribution with a Schechter function, \begin{equation} 
\phi(M)\,dM = (0.4\ln 10)\, \phi_\ast\, \mathcal{M}^{\alpha+1} e^{-\mathcal{M}} \,dM,
\label{eq:schechter} 
\end{equation} 
where $\mathcal{M}=10^{-0.4(M-M_\ast)}$, $M_\ast$ is the characteristic magnitude, $\alpha$ is the faint-end slope, and $\phi_\ast$ is a normalization that cancels in the completeness fraction. We adopt $M_\ast=-20.44$ and $\alpha=-1.3$, and integrate over $M_{\rm bright}=-24$ to $M_{\rm faint}=-12$.

The completeness of a magnitude-limited catalog is then the fraction of the luminosity function brighter than the detection threshold, \begin{equation} 
C_{\rm cat}(z,\hat{\Omega}) = \dfrac{\int_{M_{\rm bright}}^{M_{\rm thresh}(z,\hat{\Omega})} \phi(M)\,dM } {\int_{M_{\rm bright}}^{M_{\rm faint}} \phi(M)\,dM }. 
\label{eq:completeness_Cz}
\end{equation} 
Thus, $1 - C_{\rm cat}$ is the probability that the true host is fainter than the effective catalog limit at that redshift. Substituting this quantity into Eq.~\eqref{eq:Pfaint_marginal} gives the contribution of catalog incompleteness to the missing host weight.

When multiple catalogs contribute to the same transient field, we interpret $C_{\rm eff}$ as the completeness of the de-duplicated catalog union. A simple approximation is 
\begin{equation} 
C_{\rm eff}(z,\hat{\Omega}) \simeq 1 - \prod_{{\rm cat}\in\mathcal{C}_i} \left[ 1-C_{\rm cat}(z,\hat{\Omega}) \right], 
\label{eq:catalog_union_completeness} 
\end{equation} 
where $\mathcal{C}_i$ is the set of catalogs queried at the position of transient $T_i$. This assumes independent catalog incompleteness and should therefore be treated as an approximation. This completeness term is used only to set the missing host probability. It prevents the framework from assigning overly confident associations to cataloged galaxies in regions of the sky or at distance ranges where the catalog union is incomplete.

This use of the completeness term differs from the dark-siren catalog-completeness treatments of \citet{Turski2026,Borghi2026}, which marginalize catalog completeness over event populations and survey selection functions. Here the completeness is evaluated at the observed sky position of each transient and marginalized over the GW conditional redshift distribution along that line-of-sight. It therefore enters as part of the missing-host probability for an individual transient-host association, rather than as a population-level reweighting of the galaxy catalog.

\subsubsection{Hosts outside the search radius}
The second contribution in Eq.~\eqref{eq:null_weight} accounts for the possibility that the true host lies outside of the query radius,
\begin{equation}
P_{\rm out}=\int_{R_{\perp,\max}}^{\infty}p(R_\perp\midbar c)\,dR_\perp,
\label{eq:Pout}
\end{equation}
where $R_{\perp,\max} = 70\,{\rm kpc}$ and $c$ represents the source population if a class-dependent offset prior is used. Galaxies returned by the catalog query with $R_\perp>R_{\perp,\max}$ are not discarded, but they have zero offset likelihood and do not contribute to the host association posterior.

\subsubsection{Spurious or missed catalog associations}

The remaining two terms account for cases in which a cataloged object exists near the transient, but the association is unreliable. The term $P_{\rm rand}$ is the probability that the apparent host association is a chance foreground or background alignment. We estimate this from the local surface density of cataloged sources around the transient,
\begin{equation}
P_{\rm rand} = 1-\exp\left[-\pi r_{\rm assoc}^2 \sigma_{\rm bg}\right],
\label{eq:Prand}
\end{equation}
where $r_{\rm assoc}$ is an effective association radius, and $\sigma_{\rm bg}$ is the angular surface density of cataloged sources. This term is analogous to the chance-coincidence probability ($P_{\rm chance}$) used in GRB host-association studies \citep[][]{Bloom2002}, but here it enters only as a correction to the missing host weight.

The term $P_{\rm mis}$ accounts for catalog failures that are not captured by the other terms, such as a Galactic star classified as a galaxy, a deblended fragment of a larger galaxy treated as an independent source, or a spurious catalog entry. In the absence of a calibrated mis-classification model for each catalog, we treat $P_{\rm mis}$ as a small catalog-level nuisance term. We set it to zero unless a known quality flag identifies a problematic source. 

We adopt a constant value for $\pi(\varnothing)$ across all catalogs for simplicity. Both the sky- and depth-dependence of the catalog incompleteness enters through Eq.~\eqref{eq:Pfaint_marginal} instead of in the prior. Using a single value of $\pi(\varnothing)$ may assign too much missing host probability for deeper catalogs and too little for shallower or more heterogeneous catalogs. An improved approach would be to assign catalog-specific values of $\pi(\varnothing)$ calibrated to each survey's selection function, detection efficiency, depth, masking, and performance of star-galaxy separation. However, we treat the single value as a systematic limitation and leave the improved approach to future work.

\begin{figure*}[ht!]
    \centering
    \includegraphics[width=\linewidth]{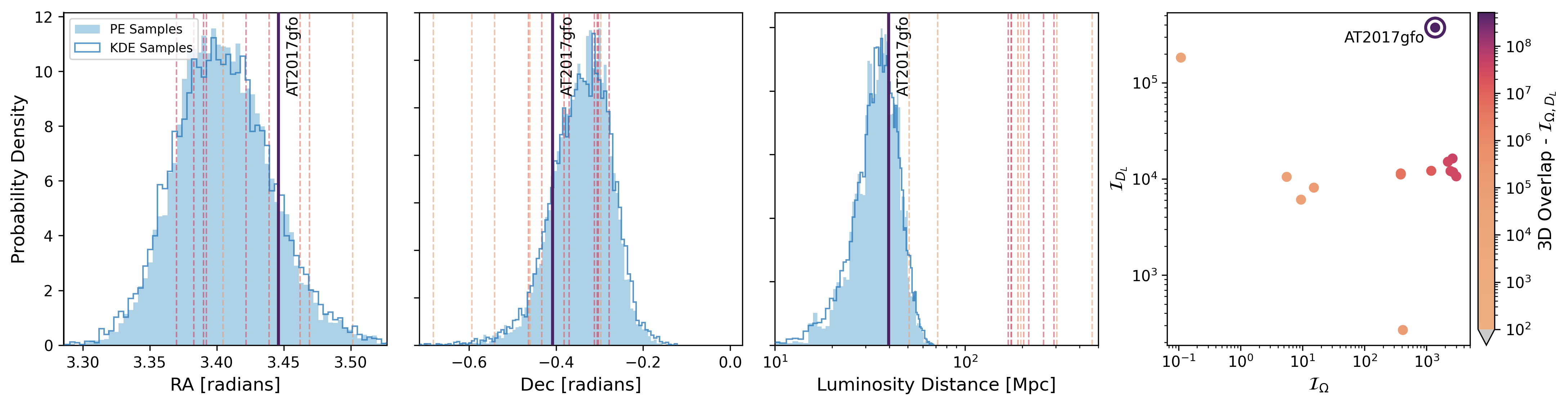}
    \caption{Marginal posterior distributions for RA (left), Declination (center) and luminosity distance $\dl$ (right) obtained from the initial probability skymap for GW170817 (\texttt{bayestar-HLV.fits.gz}). Filled histograms show the posterior samples from the \texttt{FITS} skymap and solid stepped curves are the one-dimensional (marginalized) KDE evaluations. Dashed vertical lines mark the coordinates and host-inferred distance of each EM source. The plotted sources are restricted to those within $10 \leq \dl \leq 500\,{\rm Mpc}$ and $\idl \geq 100$. Line color encodes the ranking statistic $\mathcal{B}_{\chyp/\shyp\shyp} \approx \iomegadl$ from Eq.~\eqref{eq:jointoverlap_delta}.} 
    \label{fig:margposts}
\end{figure*}

\subsection{Monte Carlo Uncertainty Propagation}
\label{sec:hostassoc:mc}

The likelihood we present above is evaluated from point estimates of the transient centroid and the galaxy centroid, morphology, and redshift. However, these measurements all carry their own non-negligible uncertainties that can broaden or skew the host association and the final GW-EM ranking statistic. We therefore propagate measurement errors with a Monte Carlo stage using a modified implementation of the Monte Carlo sampling method within the \texttt{Pr\"ost} host-association package \citep{Gagliano2025_Prost}.

For each realization $\textbf{m}=1,\dots,N_r$, we jointly perturb the transient and galaxy sky positions, the host morphology parameters $(a,b,\phi)$, and the host redshift within their reported uncertainties. For the astrometric uncertainties we assume both the transient and galaxy astrometric errors are normally distributed, and the total astrometric uncertainty follows a Rice distribution described in Sec.~\ref{sec:hostassoc:astro}. The full host association posterior is recalculated in each realization, rather than perturbing only the final statistic.

For transient $T_i$ and candidate host $\gj$, each realization gives a host-association posterior $p_\textbf{m}(\gj\midbar\dem)$, a missing host probability $p_\textbf{m}(\varnothing \midbar \dem)$, and a distance overlap for each host $\mathcal{I}_{\dl,\textbf{m},j}$. The host-marginalized distance overlap in that realization is \begin{equation} 
\mathcal{I}_{\dl,\textbf{m}} = \sum_{j=1}^{N} p_\textbf{m}(\gj\midbar \dem)\, \mathcal{I}_{\dl,j,\textbf{m}} + p_\textbf{m}(\varnothing\midbar \dem), 
\label{eq:mc_idl} 
\end{equation} 
where the final term uses $\mathcal{I}_{\dl,\textbf{m},\varnothing}=1$ for the missing host contribution. The joint $\iomegadl$ statistic is then computed within the same realization, 
\begin{equation} 
\mathcal{I}_{\Omega,\dl,\textbf{m}} = \mathcal{I}_{\Omega,\textbf{m}}\, \mathcal{I}_{\dl,\textbf{m}}. 
\label{eq:mc_iomegadl} 
\end{equation} 

The reported ranking statistic is the realization mean, 
\begin{equation} 
\left\langle \mathcal{I}_{\Omega,\dl} \right\rangle = \frac{1}{N_r} \sum_{\textbf{m}=1}^{N_r} \mathcal{I}_{\Omega,\dl,\textbf{m}},
\label{eq:mc_mean_iomegadl} 
\end{equation} 
where $N_r$ is the total number of realizations.
We compute analogous realization means for $\iomega$ and $\idl$. Because the overlap statistics are positive definite and can be skewed, we summarize the Monte Carlo scatter using the central 16th--84th percentile interval. These intervals quantify the sensitivity of the ranking to measurement uncertainties in the astrometry, morphology, and redshift observations. We report these intervals in Table~\ref{tab:gw170817stat}, but do not use them to down-rank or penalize candidates.

\begin{table*}[t]
\caption{Host galaxy associations for EM sources used in the GW170817 analysis. The transients were obtained from TNS and limited to sources which are identified within the $99\%$ probability region with no cut on discovery time. For each transient we list the TNS name; the discovery group; coordinates (RA$_{\rm EM}$, Dec$_{\rm EM}$); coordinates of the most probable host (RA$_{\rm host}$, Dec$_{\rm host}$), selected by the maximum host association probability $p_{\rm assoc}$; angular separation $\Delta\theta$ in arcseconds; projected physical offset $R_{\perp}$ in kpc; directional light radius (DLR) in arcseconds; host-normalized offset $r_{\rm DLR}$; offset likelihood $p_{\rm offset}$; host association probability $p_{\rm assoc}$, and the null probability $p_{\varnothing}$.}
\label{tab:gw170817tns}
\begin{adjustbox}{max width=\textwidth, center}
\begin{tabular}{@{}llcc@{\quad}cc@{\quad}cccc@{\quad}ccc@{}}
\toprule
Transient & Group & RA$_{\rm EM}$ & Dec$_{\rm EM}$ & RA$_{\rm host}$ & Dec$_{\rm host}$ & $\Delta\theta$ & $R_{\perp}$ & DLR & $r_{\rm DLR}$ & $p_{\rm offset}$ & $p_{\rm assoc}$ & $p_{\varnothing}$ \\
 & & ($^\circ$) & ($^\circ$) & ($^\circ$) & ($^\circ$) & ($\arcsec$) & (kpc) & ($\arcsec$) & & (\%) &(\%) & (\%) \\
\midrule
\midrule
AT2017gfo & Swope & 197.450371 & -23.381486 & 197.448776 & -23.383831 & 9.95 & 1.88 & 41.71 & 0.24 & 82.19 & 99.92 & 0.08 \\
AT2017fha & Pan-STARRS1 & 194.243435 & -17.339030 & 194.245000 & -17.342140 & 12.42 & 9.42 & 8.90 & 1.40 & 27.36 & 69.33 & 0.11 \\
SN2017djn & ATLAS & 194.358143 & -17.863531 & 194.359063 & -17.863686 & 3.20 & 4.00 & 5.33 & 0.60 & 64.15 & 99.92 & 0.07 \\
SN2017djl & ATLAS & 193.084943 & -15.960367 & 193.085342 & -15.958276 & 7.65 & 5.98 & 7.38 & 1.04 & 39.93 & 99.89 & 0.11 \\
SN2017ddy & Pan-STARRS1 & 193.816181 & -17.482037 & 193.817301 & -17.482011 & 3.85 & 4.30 & 3.49 & 1.10 & 42.56 & 97.51 & 0.10 \\
AT2017elg & GaiaAlerts & 196.054250 & -21.193600 & 196.054010 & -21.193731 & 0.93 & 0.89 & 0.95 & 0.99 & 73.97 & 99.93 & 0.07 \\
AT2016frh & GaiaAlerts & 197.041750 & -21.850239 & 197.040734 & -21.848833 & 6.09 & 4.79 & 4.49 & 1.36 & 31.29 & 99.85 & 0.15 \\
AT2017amm & Pan-STARRS1 & 195.133128 & -19.829103 & 195.135590 & -19.828634 & 8.51 & 18.77 & 7.53 & 1.13 & 36.32 & 99.87 & 0.13 \\
AT2017emn & Pan-STARRS1 & 195.133023 & -20.061578 & 195.133461 & -20.061772 & 1.64 & 4.92 & 3.80 & 0.43 & 79.21 & 99.94 & 0.06 \\
AT2017ejv & Pan-STARRS1 & 198.371810 & -26.572973 & 198.371777 & -26.572467 & 1.82 & 1.64 & 1.65 & 1.11 & 54.22 & 96.83 & 0.08 \\
SN2017bzm & ATLAS & 198.767330 & -24.801523 & 198.768000 & -24.800690 & 3.71 & 3.13 & 10.53 & 0.35 & 76.19 & 99.93 & 0.07 \\
AT2017cvf & Pan-STARRS1 & 194.108003 & -15.874554 & 194.108932 & -15.874316 & 3.33 & 7.63 & 3.96 & 0.84 & 53.40 & 99.91 & 0.09 \\
AT2017aor & Pan-STARRS1 & 195.441116 & -22.449582 & 195.440252 & -22.449528 & 2.88 & 6.11 & 3.59 & 0.80 & 56.59 & 99.92 & 0.08 \\
AT2017app & Pan-STARRS1 & 193.425959 & -15.045208 & 193.426078 & -15.045525 & 1.21 & 4.22 & 2.74 & 0.44 & 82.55 & 99.95 & 0.05 \\
AT2017dsr & Pan-STARRS1 & 194.997441 & -20.223033 & 194.997272 & -20.222823 & 0.95 & 6.42 & 2.87 & 0.33 & 88.93 & 99.95 & 0.05 \\
AT2017emw & Pan-STARRS1 & 196.751970 & -19.738653 & 196.750757 & -19.739330 & 4.78 & 14.37 & 2.15 & 2.22 & 16.47 & 99.73 & 0.26 \\
AT2017aoq & Pan-STARRS1 & 198.237002 & -22.494327 & 198.236975 & -22.494400 & 0.28 & 0.94 & 0.46 & 0.60 & 95.59 & 99.81 & 0.06 \\
AT2017esn & GaiaAlerts & 195.974375 & -19.162839 & 195.975906 & -19.159542 & 12.96 & 10.94 & 23.24 & 0.56 & 59.85 & 93.78 & 0.07 \\
AT2017aod & Pan-STARRS1 & 202.390557 & -31.000348 & 202.390818 & -31.000636 & 1.31 & 1.70 & 3.26 & 0.40 & 82.96 & 99.94 & 0.06 \\
AT2017esx & GaiaAlerts & 195.068292 & -17.027861 & 195.067978 & -17.027241 & 2.48 & 0.83 & 14.19 & 0.17 & 89.02 & 99.95 & 0.05 \\
AT2017bej & ATLAS & 202.714896 & -31.760956 & 202.715579 & -31.761871 & 3.90 & 9.49 & 1.95 & 2.00 & 21.44 & 73.23 & 0.17 \\
AT2016ity & GaiaAlerts & 204.885833 & -39.068300 & 204.888382 & -39.071697 & 14.15 & 12.35 & 5.13 & 2.76 & 7.53 & 99.20 & 0.80 \\
AT2016fru & GaiaAlerts & 200.619583 & -26.381961 & 200.619230 & -26.382736 & 3.01 & 5.60 & 6.25 & 0.48 & 70.48 & 99.93 & 0.06 \\
SN2017dps & ASAS-SN & 204.166829 & -33.967025 & 204.162720 & -33.965916 & 12.90 & 3.10 & 113.27 & 0.11 & 100.00 & 99.94 & 0.06 \\
AT2017cyu & SkyMapper & 202.996675 & -31.766430 & 202.996565 & -31.766315 & 0.53 & 2.34 & 1.05 & 0.51 & 90.98 & 87.02 & 0.05 \\
AT2017anr & Pan-STARRS1 & 195.509215 & -18.047855 & 195.509035 & -18.047556 & 1.24 & --- & 0.79 & 1.56 & 55.48 & 99.90 & 0.09 \\
AT2017amn & Pan-STARRS1 & 194.842791 & -16.166754 & 194.841256 & -16.166755 & 5.31 & 31.47 & 0.18 & 29.83 & 0.00 & 0.00 & 100.00 \\
AT2017arb & Pan-STARRS1 & 190.588954 & -10.923021 & 190.595997 & -10.929809 & 34.89 & --- & 2.23 & 15.67 & 0.00 & 0.05 & 99.95 \\
AT2017fpo & GaiaAlerts & 206.003042 & -40.478950 & 206.011219 & -40.527060 & 174.64 & 420.85 & 3.41 & 51.27 & 0.00 & 0.00 & 100.00 \\
AT2016iuu & GaiaAlerts & 214.538042 & -54.598411 & 214.548676 & -54.656143 & 209.01 & 337.43 & 6.56 & 31.87 & 0.00 & 0.00 & 100.00 \\
AT2016iuo & GaiaAlerts & 212.355708 & -55.298431 & 212.171890 & -55.235096 & 440.61 & 123.11 & 25.37 & 17.38 & 0.00 & 0.00 & 100.00 \\
AT2016hst & GaiaAlerts & 219.934083 & -60.927731 & --- & --- & --- & --- & --- & --- & --- & 0.00 & 100.00 \\
AT2016hrh & GaiaAlerts & 218.757542 & -57.762489 & --- & --- & --- & --- & --- & --- & --- & 0.00 & 100.00 \\

\bottomrule
\end{tabular}
\end{adjustbox}
\end{table*}

\begin{table*}[t]
\centering
\caption{EM sources used in the GW170817 analysis ranked by $\iomegadl$, Eq.~\eqref{eq:jointoverlap_delta}, from the initial skymap \texttt{bayestar-HLV.fits.gz}. For each transient we report the TNS candidate name; redshift $z$ (cosmological frame, with peculiar velocity broadening) and its uncertainty $\sigma_z$; redshift type as spectroscopic (spec), photometric (phot), or None (-); the catalog providing the final $z$; luminosity distance $\dl$ and $\sigma_{\dl}$ in Mpc, assuming a flat $\Lambda$CDM cosmology; and the Monte Carlo means for $\iomega$, $\idl$, and $\iomegadl$. The quantity $\Delta\iomegadl$ is the offsets from the mean to the 16th and 84th percentiles across Monte Carlo realizations. Missing redshifts are denoted by --- and the EM distance posterior is taken to be the (uninformative) uniform in comoving volume prior. The highest-ranked entry is AT2017gfo, the confirmed counterpart to GW170817. The sources AT2016hst and AT2016hrh were not found in any of the catalogs. None of the pre-merger sources have $\iomegadl$ greater than or equal to that of AT2017gfo and the p-values from Eqs.~(\ref{eq:pvalue_bg},\ref{eq:pvalue_min}) are $p_{\rm cand}=0.030$ and $p_{\rm min}=0.030$. Because AT2017gfo is the only post-merger candidate, there is no multiple-candidate correction applied and $p_{\rm event}=0.030$ from Eq.~\eqref{eq:pvalue_event}.
}
\label{tab:gw170817stat}
\begin{adjustbox}{max width=\textwidth, center}
\begin{tabular}{@{}llccc@{\quad}cc@{\quad}c@{\quad}c@{\quad}c@{\quad}l}
\toprule
Transient & Catalog & $z$ & $\sigma_z$ & Type & $\dl$ & $\sigma_{\dl}$ & $\langle\iomega\rangle$ & $\langle\idl\rangle$ & $\langle\iomegadl\rangle$ & \multicolumn{1}{c}{$\Delta\iomegadl$} \\
\midrule
AT2017gfo & GLADE+ & $0.0092$ & $0.0004$ & spec & $39.64$ & $1.59$ & $1.38 \times 10^{3}$ & $3.77\times10^{5}$ & $5.19\times10^{8}$ & $[-0.09,\,+0.07]\times10^{8}$ \\
AT2017fha & GLADE+ & $0.0383$ & $0.0003$ & phot & $168.61$ & $1.38$ & $2.62 \times 10^{3}$ & $1.69\times10^{4}$ & $4.43\times10^{7}$ & $[-1.02,\,+0.96]\times10^{7}$ \\
SN2017djn & GLADE+ & $0.0651$ & $0.0351$ & phot & $292.60$ & $168.79$ & $3.01 \times 10^{3}$ & $1.14\times10^{4}$ & $3.45\times10^{7}$ & $[-1.24,\,+1.49]\times10^{7}$ \\
SN2017djl & GLADE+ & $0.0395$ & $0.0003$ & phot & $174.15$ & $1.54$ & $2.21 \times 10^{3}$ & $1.51\times10^{4}$ & $3.33\times10^{7}$ & $[-0.01,\,+0.01]\times10^{7}$ \\
SN2017ddy & LS-DR10 & $0.0578$ & $0.0158$ & phot & $258.20$ & $74.29$ & $2.68 \times 10^{3}$ & $1.16\times10^{4}$ & $3.10\times10^{7}$ & $[-0.54,\,+0.43]\times10^{7}$ \\
AT2017elg & LS-DR10 & $0.0485$ & $0.0317$ & phot & $215.50$ & $148.81$ & $2.43 \times 10^{3}$ & $1.14\times10^{4}$ & $2.78\times10^{7}$ & $[-1.01,\,+1.28]\times10^{7}$ \\
AT2016frh & GLADE+ & $0.0397$ & $0.0004$ & phot & $175.15$ & $1.85$ & $1.19 \times 10^{3}$ & $1.21\times10^{4}$ & $1.44\times10^{7}$ & $[-0.01,\,+0.01]\times10^{7}$ \\
AT2017amm & GLADE+ & $0.1228$ & $0.0150$ & phot & $573.88$ & $76.33$ & $2.62 \times 10^{3}$ & $4.66\times10^{3}$ & $1.22\times10^{7}$ & $[-0.24,\,+0.25]\times10^{7}$ \\
AT2017emn & PS1-DR2 & $0.1778$ & $0.0373$ & phot & $860.15$ & $203.23$ & $2.36 \times 10^{3}$ & $2.12\times10^{3}$ & $5.01\times10^{6}$ & $[-2.41,\,+2.37]\times10^{6}$ \\
AT2017ejv & LS-DR10 & $0.0459$ & $0.0214$ & phot & $203.25$ & $99.22$ & $3.84\times 10^{2}$ & $1.12\times10^{4}$ & $4.31\times10^{6}$ & $[-1.17,\,+0.93]\times10^{6}$ \\
SN2017bzm & GLADE+ & $0.0428$ & $0.0010$ & phot & $189.29$ & $4.62$ & $3.83\times 10^{2}$ & $1.12\times10^{4}$ & $4.27\times10^{6}$ & $[-0.04,\,+0.04]\times10^{6}$ \\
AT2017cvf & GLADE+ & $0.1284$ & $0.0372$ & phot & $602.22$ & $192.81$ & $7.36\times 10^{2}$ & $5.30\times10^{3}$ & $3.90\times10^{6}$ & $[-1.71,\,+1.88]\times10^{6}$ \\
AT2017aor & GLADE+ & $0.1173$ & $0.0368$ & phot & $546.10$ & $188.56$ & $3.47\times 10^{2}$ & $6.04\times10^{3}$ & $2.09\times10^{6}$ & $[-0.99,\,+1.02]\times10^{6}$ \\
AT2017app & GLADE+ & $0.2141$ & $0.0400$ & phot & $1057.97$ & $226.11$ & $9.28\times 10^{2}$ & $1.45\times10^{3}$ & $1.34\times10^{6}$ & $[-0.89,\,+0.66]\times10^{6}$ \\
AT2017dsr & PS1-DR2 & $0.6193$ & $0.1980$ & phot & $3668.37$ & $1482.85$ & $1.83\times 10^{3}$ & $4.14\times 10^{2}$ & $7.57\times10^{5}$ & $[-7.02,\,+6.15]\times10^{5}$ \\
AT2017emw & GLADE+ & $0.1779$ & $0.0388$ & phot & $860.61$ & $211.96$ & $1.57\times 10^{2}$ & $2.04\times10^{3}$ & $3.20\times10^{5}$ & $[-1.60,\,+1.56]\times10^{5}$ \\
AT2017aoq & LS-DR10 & $0.2060$ & $0.0978$ & phot & $1013.08$ & $562.42$ & $79.43$ & $3.00\times10^{3}$ & $2.38\times10^{5}$ & $[-2.16,\,+2.85]\times10^{5}$ \\
AT2017esn & GLADE+ & $0.0428$ & $0.0007$ & spec & $189.34$ & $3.34$ & $8.03\times 10^{2}$ & $2.01\times 10^{2}$ & $1.62\times10^{5}$ & $[-1.62,\,+2.44]\times10^{5}$ \\
AT2017aod & GLADE+ & $0.0675$ & $0.0352$ & phot & $303.93$ & $169.72$ & $15.24$ & $7.55\times10^{3}$ & $1.15\times10^{5}$ & $[-0.45,\,+0.59]\times10^{5}$ \\
AT2017esx & GLADE+ & $0.0165$ & $0.0004$ & spec & $71.57$ & $1.87$ & $4.15\times 10^{2}$ & $2.29\times 10^{2}$ & $9.49\times10^{4}$ & $[-5.43,\,+7.37]\times10^{4}$ \\
AT2017bej & LS-DR10 & $0.1377$ & $0.0743$ & phot & $649.55$ & $396.29$ & $10.70$ & $6.24\times10^{3}$ & $6.67\times10^{4}$ & $[-2.76,\,+3.43]\times10^{4}$ \\
AT2016ity & GLADE+ & $0.0443$ & $0.0004$ & phot & $196.26$ & $1.75$ & $5.55$ & $1.05\times10^{4}$ & $5.83\times10^{4}$ & $[-0.02,\,+0.02]\times10^{4}$ \\
AT2016fru & GLADE+ & $0.1008$ & $0.0363$ & phot & $464.46$ & $182.30$ & $9.46$ & $5.83\times10^{3}$ & $5.52\times10^{4}$ & $[-2.70,\,+2.67]\times10^{4}$ \\
SN2017dps & GLADE+ & $0.0117$ & $0.0008$ & spec & $50.73$ & $3.30$ & $0.11$ & $1.78\times10^{5}$ & $1.95\times10^{4}$ & $[-0.30,\,+0.38]\times10^{4}$ \\
AT2017cyu & LS-DR10 & $0.2928$ & $0.1316$ & phot & $1509.59$ & $820.03$ & $6.09$ & $1.24\times10^{3}$ & $7.58\times10^{3}$ & $[-6.92,\,+4.42]\times10^{3}$ \\
AT2017anr & LS-DR10 & --- & --- & --- & --- & --- & $6.40\times 10^{2}$ & $1.00$ & $6.40\times 10^{2}$ & $[-0.00,\,+0.00]\times10^{2}$ \\
AT2017amn & LS-DR10 & $0.4742$ & $0.0943$ & phot & $2658.56$ & $648.39$ & $1.23\times 10^{2}$ & $1.02$ & $1.25\times10^{2}$ & $[-0.03,\,-0.01]\times10^{2}$ \\
AT2017arb & PS1-DR2 & --- & --- & --- & --- & --- & $1.01\times 10^{2}$ & $1.00$ & $1.01\times10^{2}$ & $[-0.00,\,+0.00]\times10^{2}$ \\
AT2017fpo & GLADE+ & $0.1361$ & $0.0375$ & phot & $641.63$ & $195.83$ & $2.41$ & $4.71$ & $1.14\times10^{1}$ & $[-0.90,\,+2.73]\times10^{1}$ \\
AT2016iuu & GLADE+ & $0.0861$ & $0.0150$ & phot & $392.84$ & $73.11$ & $4.57$ & $1.00$ & $4.57$ & $[-0.01,\,+0.01]$ \\
AT2016iuo & GLADE+ & $0.0137$ & $0.0007$ & spec & $59.22$ & $3.08$ & $0.66$ & $1.00$ & $0.66$ & $[-0.003,\,+0.003]$ \\
AT2016hst & --- & --- & --- & --- & --- & --- & $\ll 1$ & $1.00$ & $\ll 1$ & $\ll 1$ \\
AT2016hrh & --- & --- & --- & --- & --- & --- & $\ll 1$ & $1.00$ & $\ll 1$ & $\ll 1$ \\\\
\bottomrule
\end{tabular}
\end{adjustbox}
\end{table*}

\section{Background estimation and association significance}
\label{sec:significance}
The ranking statistic developed in Secs.~\ref{sec:bayes} and ~\ref{sec:hostassoc} provides a relative ordering of the EM candidates based on their agreement with the GW localization, luminosity distance and host galaxy information. The ordering on its own does not determine the significance of an association. We estimate the significance of a candidate using an empirical p-value by comparing the ranking value with the distribution of ranking values for sources known to be unrelated to the GW event. This is analogous to the approach used to estimate the empirical background to calibrate ranking statistics in GW searches \citep[e.g.,]{Capano2016}.

The ranking statistic considered here is $\iomegadl$. It can be extended to include additional information, such as a temporal overlap term $\mathcal{I}_{t}$ based on the time of first detection, an inclination-angle overlap term $\mathcal{I}_{\iota}$ \citep{Piotrzkowski2022}, or using terms that describe the transient's properties, such as peak magnitude or lightcurve evolution. While these additional terms can lead to further separation between candidates or change the relative ordering of candidates, they do not enter into the significance calculation here, as it is based only on $\iomegadl$.

As the number of potential candidates increases, so does the probability that an unrelated transient will have a large value of $\iomegadl$ by chance. This is particularly relevant for Rubin \citep{Ivezic2019}, where the transient alert rate, and thus the number of unrelated candidates associated with a GW event, is expected to be orders of magnitude larger than for current optical wide-field surveys \citep{Moller2021, Stevenson2026}.

\subsection{Association p-value}
\label{sec:pvalue}

Transients discovered before the GW trigger time cannot originate from the subsequent GW event, and thus, provide an empirical population of unrelated astrophysical transients under the unrelated-source hypothesis $\hhyp_{\shyp\shyp}$. We use the pre-merger TNS sources described in Sec.~\ref{sec:tns_candidates}, restricted to the same 99\% GW localization region and to the year preceding the GW trigger time, to generate the background distribution. Each background source is processed through the same host-association and ranking process as the post-merger candidates. 

The transient classification reported by TNS does not enter the calculation of $\iomegadl$. Thus, we use all optical transients in the pre-merger sample, regardless of their reported astrophysical classification. If the ranking statistic is extended to include classification, lightcurve, or other transient-dependent information, that same information and selection criteria must be included when constructing the background distribution.

For candidate $T_i$, we refer to the ranking statistic as $\mathcal{I}_{i} \equiv \mathcal{I}_{\Omega, \dl,i}$ and define $N_{\rm bg}$ as the number of pre-merger background sources. The p-value for candidate $T_i$ is
\begin{equation}
p_{{\rm cand},i} = \frac{k_i+1}{N_{\rm bg}+1},
\label{eq:pvalue_bg}
\end{equation}
where 
\begin{equation}
k_i = \sum_{m=1}^{N_{\rm bg}} \mathbb{1} \left[\mathcal{I}_{m} \geq \mathcal{I}_{i} \right],
\label{eq:k_bg}
\end{equation}
is the number of background sources with ranking statistics equal to or greater than the candidate and $\mathbb{1}[ \, ]$ is the indicator function. The $+1$ terms in Eq.~\eqref{eq:pvalue_bg} avoid returning a p-value of zero when there are no background sources exceeding the candidate in ranking \citep{North2002, Phipson2010}. 
If $k_i=0$, then the minimum p-value that can be measured from a background of $N_{\rm bg}$ sources is
\begin{equation}
p_{\rm min}=\frac{1}{N_{\rm bg}+1}.
\label{eq:pvalue_min}
\end{equation}
This tells us that if the candidate exceeds every value in the observed background, then any further increase in $\iomegadl$ would not lead to a decrease in $p_{\rm cand}$. A smaller p-value can only be generated through a larger sample of background sources.

Equation~\eqref{eq:pvalue_bg} assumes that under $\hhyp_{\shyp\shyp}$ the distribution of $\iomegadl$ from the pre-merger sources is representative of that for the unrelated post-merger sources.
We refer to this as an \textit{exchangeability} assumption \citep[see e.g.,][for its use in an alternative statistical treatment of GW searches]{Ashton2024}. However, this assumption may not be exact for two main reasons. First, TNS combines discoveries from multiple surveys and follow-up programs, and does not correspond to a single, uniform survey selection function. In addition, even for the same telescope, the post-merger candidates can be discovered from targeted observations that are initiated in response to the GW alert. The targeted observations may choose to increase the cadence or exposure time, or may preferentially cover regions of high GW probability or target galaxies consistent with the GW distance, all of which can increase the number of candidates within portions of the localization region, relative to the number reported otherwise. These differences can alter the distribution of $\iomegadl$ for unrelated post-merger candidates. Thus, the p-values reported here are conditional on the TNS background population used in this analysis and are not survey-independent probabilities.

\subsection{Low-number backgrounds}
\label{sec:low_background}
For events with compact GW localizations, only a small number of pre-merger sources may be available for estimating the background distribution. If a candidate has a ranking statistic larger than all $N_{\rm bg}$ background sources, Eq.~\eqref{eq:pvalue_bg} will quantify the relative ordering compared to the background, but is ultimately limited to $p_{\rm min}$ from Eq.~\eqref{eq:pvalue_min}.
This limitation is relevant for GW170817, where there are only $N_{\rm bg}=32$ sources available, and $p_{\rm min}=0.030$.

With only a small number of background sources, the p-value can establish that AT2017gfo ranks above all $N_{\rm bg}=32$ sources, but it cannot distinguish how far the candidate is above the background once the limit of $p_{\rm min}$ is reached. To determine whether combinations of the observed background properties and positions can produce ranking values closer to or as comparably large as that of the candidate, we perform what we refer to here as a \textit{cross-pair calculation}.

For each pre-merger source $T_m$, we define $E_m$ as the EM and host galaxy information, including the transient-host offset and orientation, host galaxy morphology, and redshift. We evaluate $E_m$ at the observed sky position $\hat{\Omega}_n$ of every other background source $T_n$, where $m\neq n$. At each position, we re-evaluate the GW sky probability and line-of-sight distance posterior, catalog completeness, and re-normalize the host-association and missing host probabilities before recalculating $\iomegadl$.
For $N_{\times,\rm bg}=30$ sources, this produces
\begin{equation}
N_{\times}=N_{\times,\rm bg}(N_{\times,\rm bg}-1)
\end{equation}
ordered cross-pair values with ranking statistics 
\begin{equation}
\mathcal{I}_{mn} \equiv \mathcal{I}(E_m, \hat\Omega_n), \qquad m\neq n.
\end{equation}
We use the observed positions rather than drawing random positions uniformly over the GW localization so that the calculation preserves the spatial sampling of the TNS background.

Within statistical literature, reassigning observed properties among fixed observed locations is related to the random-labeling construction used for marked spatial point patterns \citep{FeigelsonBabu2012, Baddeley2015}.
Random-labeling refers to reassigning observed properties (marks) among a fixed set of positions (spatial points) and is a method used to test the independence of marks with the positions \citep{Baddeley2015}. The cross-pair calculation is not a formal random-labeling test. Instead, we use the reassignment principle to test whether alternative pairings of the properties $E_m$ and the fixed source positions $\hat\Omega_n$ can produce ranking statistic values comparable to that of the candidate. It is meant only as an additional robustness check and not to introduce additional independent background events. 

However, performing the cross-pair calculation introduces an additional exchangeability assumption, i.e., that the properties $E_m$ measured for one background source do not vary strongly with the sky position used in the pairing. The assumption may not hold exactly if properties of the detected transient population vary across the sky. For example, a deeper survey may report more distant transients or better-characterized host galaxies than a shallower survey, or sky-dependent variations in galaxy catalogs may lead to significantly different host association probabilities. Thus, some cross-pair combinations may not be representative of the actual observations.

Although the cross-pair calculation produces $N_{\times}$ ranking statistic values, these are not $N_{\times}$ independent background events. The same $N_{\times,\rm bg}$ sets of $E_m$ and the same $N_{\times,\rm bg}$ observed sky positions $\hat\Omega_{n}$ are reused in multiple combinations. A similar limitation occurs in "time-slide" background estimation for GW searches, where it was found that increasing the number of time slides to repeatedly recombine a finite set of GW triggers does not lead to unlimited increase in independent background information \citep{Was2009}\citep[see also][]{Capano2016}.

Thus, the $N_{\times}$ cross-pairs are only used as an additional comparison with AT2017gfo and not to reduce the empirical p-value below the floor of $p_{\rm min}$ in Eq.~\eqref{eq:pvalue_min}.

\subsection{Multiple candidates per event}
\label{sec:trials}
The p-value in Eq.~\eqref{eq:pvalue_bg} is the significance of an \textit{individual} candidate. For a GW event with multiple ($N_{\rm cand}>1$) post-merger candidates, there is an increased probability that at least one unrelated source will have a small p-value by chance. This introduces a trials factor, analogous to the look-elsewhere effect that arises when assessing significance across multiple trials \citep{Lyons2008, GrossVitells2010}.

We account for this by treating the candidate associations as a family of hypothesis tests and applying the Holm-Bonferroni (or Holm) method \citep{Holm1979}. The method orders the p-values as
\begin{equation}
p_{{\rm cand},(1)} \leq p_{{\rm cand},(2)} \leq \cdots \leq p_{{\rm cand},(N_{\rm cand})},
\end{equation}
where $N_{\rm cand}$ is the total number of post-merger candidates tested for a given GW event. The Holm adjusted p-value of the candidate at ordered rank $(i)$ is 
\begin{equation}
p_{{\rm corr},(i)} = \min\left[1, \, \max_{j\leq i} \left\{ \left(N_{\rm cand}-j+1\right) p_{{\rm cand},(j)} \right\} \right].
\label{eq:pvalue_holm}
\end{equation}
If a significance threshold $\hat\alpha$ is applied to the corrected p-values, then, under the hypothesis that all of the candidates are unrelated, the probability that any candidate is found to be significant by chance is at most $\hat\alpha$. This result does not require the individual candidate tests to be statistically independent. However, it does assume that the pre-merger background is representative of the ranking statistics expected for unrelated post-merger candidates. 

For the candidate with the smallest p-value $p_{\rm cand}$, Eq.~\eqref{eq:pvalue_holm} reduces to 
\begin{equation}
p_{{\rm corr},(1)} = \min \left[ 1, N_{\rm cand}\,p_{{\rm cand},(1)} \right] \equiv p_{\rm event}.
\label{eq:pvalue_event}
\end{equation}
We define $p_{\rm event}$ as the Holm corrected p-value after accounting for the number of post-merger candidates. We note that $p_{\rm event}$ should not be interpreted as a calibrated event false-alarm probability.

\begin{table*}[t]
\centering
\caption{Top 5 candidate counterparts to GW190425 ranked by $\langle\iomegadl \rangle$ and their host galaxy identifications. The candidates were obtained from TNS and limited to sources which are identified within the 99\% probability region and discovered after the GW event time.
For each candidate transient we report the TNS candidate name; discovery group, coordinates (RA$_{\rm EM}$, Dec$_{\rm EM}$); coordinates of the most probable host (RA$_{\rm host}$, Dec$_{\rm host}$); host-normalized offset $r_{\rm DLR}$; host association probability $p_{\rm assoc}$; redshift $z$ (cosmological-frame, with peculiar velocity broadening) and uncertainty $\sigma_z$; luminosity distance $\dl$ and uncertainty $\sigma_{\dl}$ in Mpc and assuming a flat $\Lambda$CDM cosmology; the Monte Carlo means for $\iomega$, $\idl$, and $\iomegadl$; the p-values calculated from the pre-merger sources $p_{\rm cand}$ using Eq.~\eqref{eq:pvalue_bg}; and the Holm-corrected p-values $p_{\rm corr}$. 
AT2019dzk has the smallest p-value of $p_{\rm cand}=5.88\times10^{-4}$. After correcting for all the 433 candidates, the p-value is Holm-corrected to $p_{\rm corr}=0.255$. As this is the smallest corrected p-value for the event, $p_{\rm event}=0.255$.}
\label{tab:gw190425}
\begin{adjustbox}{max width=\textwidth, center}
\begin{tabular}{@{}ll@{\quad}cc@{\quad}cc@{\quad}c@{\quad}c@{\quad}c@{\quad}c@{\quad}c@{\quad}c@{\quad}c@{\quad}c@{\quad}c@{}}
\toprule
Transient & Group & RA$_{\rm EM}$ & Dec$_{\rm EM}$ & RA$_{\rm host}$ & Dec$_{\rm host}$ & $r_{\rm DLR}$ & $p_{\rm assoc}$ & $z \pm \sigma_{z}$ & $\dl \pm \sigma_{\dl}$ & $\langle\iomega\rangle$ & $\langle\idl\rangle$ & $\langle\iomegadl\rangle$ 
& $p_{\rm cand}$ & $p_{\rm corr}$\\
& & ($^\circ$) & ($^\circ$) & ($^\circ$) & ($^\circ$) & & (\%) & & (Mpc) & & & 
& & \\
\midrule
\midrule
AT2019dzk & ZTF & 258.341454 & -9.964466 & 258.338074 & -9.964910 & 0.75 & 74.35 & $0.023 \pm 0.001$ & $102.13 \pm 2.65$ & 3.46 & 7038 & 24334 & $5.88\times10^{-4}$ & 0.255 \\
SN2019dzw & ZTF & 262.791487 & -8.450722 & 262.792389 & -8.452872 & 1.78 & 50.77 & $0.028 \pm 0.001$ & $122.76 \pm 3.65$ & 3.35 & 5471 & 18351 & $1.37\times10^{-3}$ & 0.593 \\
AT2019eit & Gaia & 20.702667 & -38.787411 & 20.704187 & -38.785809 & 1.17 & 99.05 & $0.033 \pm 0.001$ & $143.98 \pm 2.36$ & 2.32 & 7584 & 17568 & $1.37\times10^{-3}$ & 0.593 \\
AT2019esn & SNHunt & 222.983831 & 51.264207 & 222.988907 & 51.264881 & 1.13 & 80.78 & $0.026 \pm 0.000$ & $111.85 \pm 1.92$ & 3.55 & 4667 & 16582 & $1.57\times10^{-3}$ & 0.674 \\
AT2019ewb & ZTF & 221.652363 & 56.234219 & 221.654205 & 56.233078 & 0.50 & 88.06 & $0.039 \pm 0.001$ & $171.91 \pm 2.72$ & 3.03 & 3956 & 11981 & $7.84\times10^{-3}$ & 1.000 \\
\bottomrule
\end{tabular}
\end{adjustbox}
\end{table*}

\begin{figure*}[t]
    \centering
    \includegraphics[width=\linewidth]{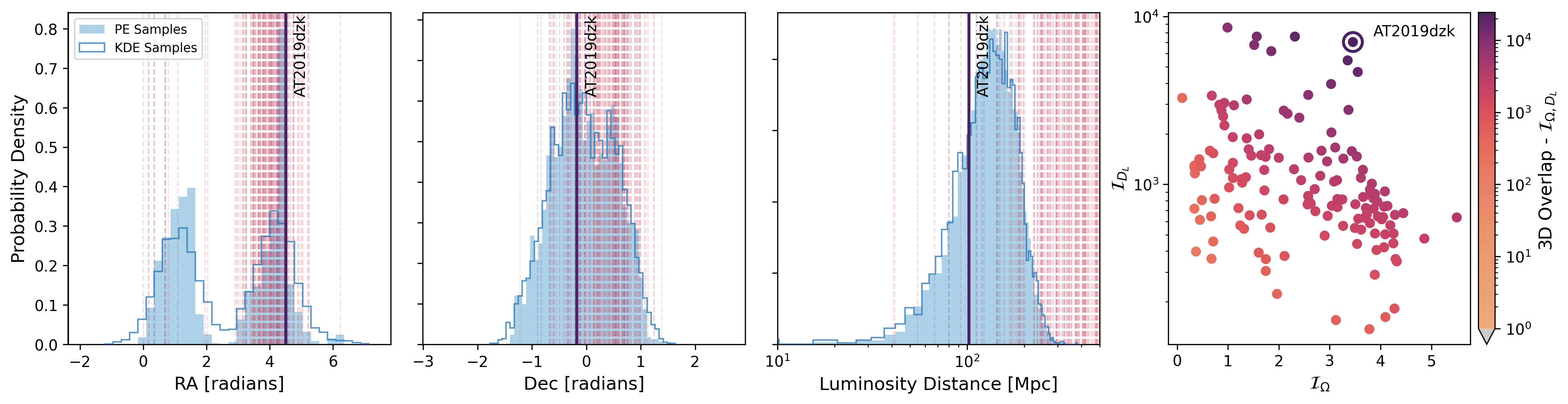}
    \caption{Marginal posterior distributions for RA (left), Declination (center), and luminosity distance $\dl$ (right) for GW190425 using the GWTC-3.0 skymap. Dashed vertical lines mark the coordinates and host-inferred distances of the plotted EM candidates. The plotted candidate set is restricted to $10\leq\dl\leq500\,{\rm Mpc}$, $\idl\geq100$, and $\iomega\geq0.1$. Line color encodes the ranking statistic $\mathcal{B}_{\chyp/\shyp\shyp}\approx\iomegadl$ from Eq.~\eqref{eq:jointoverlap_delta}.}
    \label{fig:gw190425_margposts}
\end{figure*}

\section{Results}
\label{sec:results}

We apply the ranking statistic of Eq.~\eqref{eq:jointoverlap_delta} to transients reported to TNS for two BNS mergers, GW170817 and GW190425. GW170817 is a high signal-to-noise ratio event localized to $\sim$28~deg$^2$ by a three-detector network with a securely identified optical counterpart at $\sim$40~Mpc, well within the GLADE+ completeness limit \citep{Dalya2022}.

GW190425 is the second BNS merger event \citep{GW190425}. Only two detectors were online, which produced a broad, multi-modal sky localization of $\sim$8284~deg$^2$ with an inferred luminosity distance of $159^{+69}_{-71}$~Mpc \citep{GW190425}. Despite extensive follow-up, no confident EM counterpart was ever identified \citep[e.g.,][]{Antier2020, Boersma2021, Chang2021, Coughlin2019, Coulter2025, Gompertz2020, Hosseinzadeh2019,Keinan2026, Lundquist2019,Mo2023, Paek2024, Paterson2021, Pozanenko2020, Raman2025, SagusCarracedo2021, Sasada2021, Smartt2024}.

Thus, GW170817 is a best-case scenario and is used to validate that the statistic recovers the known counterpart while the GW190425 example tests how the statistic behaves when the data do not support a confident association.

For each event, we re-weight the skymap to a common luminosity distance prior, or uniform in comoving volume on fixed bounds $\dl\in[10,10^4]$\,Mpc, so that $\idl$ is comparable across candidates and events.

For host association we use the host-normalized DLR offset likelihood, $\rdlr$, given by a Rice astrometric kernel convolved with a Gamma prior for the intrinsic offsets (Sec.~\ref{sec:hostassoc}). For the Gamma prior we adopt $k_c=2$ and $\lambda_c=1$ as our baseline. Based on observed short GRB offsets from \citet{Fong2022}, we find $\lambda_c=0.87^{+0.36}_{-0.13}$. We round to $\lambda_c=1$ for simplicity and to avoid over-tuning to the small calibration set. Our results are insensitive to this choice within the quoted uncertainty. We take transient-centroid uncertainties of $\sigma_{\rm EM}\!\approx\!0\farcs5$, host-centroid errors from LS-DR10 ($\sigma\!\approx\!0\farcs1$), and a registration term of $\sigma_{\rm tie}\!=\!0\farcs2$.

\subsection{GW170817}
\label{sec:gw170817}

We use the initial low-latency BAYESTAR skymap\footnote{\texttt{bayestar-HLV.fits.gz}, see \url{https://gracedb.ligo.org/events/G298048}} rather than the GWTC posterior samples \citep{GWTC3} because the latter data are conditioned on the known counterpart location and host distance as priors. Using the sample would be circular evaluation of the framework, since it would rank the counterpart against itself. We find 33 spatially coincident sources within the 99\% probability contour reported to TNS. All but one (AT2017gfo) were discovered prior to the GW trigger time and are used as the unrelated-source background. All of the sources are processed as if they were potential candidates, and thus, the ranking of AT2017gfo can be compared directly with the pre-merger background sources. 

We show the host association matches with their probabilities in Table~\ref{tab:gw170817tns} and the transient rankings in Table~\ref{tab:gw170817stat}. Over the set of TNS sources inside the 99\% GW skymap contour, AT2017gfo ranks first by a factor of $\sim 12$ above the highest-ranked background source. The host of AT2017gfo is correctly recovered (NGC\,4993) with a high association probability $p_{\rm assoc}\sim 99.93\%$ as shown in Fig.~\ref{fig:hostassoc_dlr}. At the position of AT2017gfo and distance of NGC\,4993 we find $\iomega\approx 1.38\times 10^3$, $\idl\approx 3.77\times 10^5$, and $\iomegadl\approx 5.19\times 10^8$.

For the host galaxy NGC\,4993 we obtain the heliocentric $\zspec=0.0092\pm0.0004$ in GLADE+. LS-DR10 and PS1-STRM report discordant values of $\zphot\!= 0.317 \,(0.076) \pm 0.855$ for LS-DR10 mean (median) and an extrapolated Monte Carlo estimate of $\zphot\!=\!0.089\pm0.015$ in PS1-STRM. These discrepancies are reflective of known low-$z$ systematics \citep{Beck2017, Bolzonella2000}. The host galaxy coordinates were also crossmatched against NED which reports $\zphot\!=\!0.00982 \pm 0.00002$ \citep{Pan2017} with a redshift flag of \texttt{UUN}. This translates to ``unknown technique and uncertain method" and would not be used to over-write the cataloged GLADE+ value.

In Fig.~\ref{fig:dls_gw170817} we show the GW line-of-sight distance posterior $p(\dl\midbar\hat\Omega, \dgw)$, the EM mixture components mapped from the host photometric redshift, the resulting EM posterior $p(\dl\midbar\gj,\dem)$, and the product proportional to the distance-overlap $\idl$ integrand for AT2017gfo.

The next highest ranked candidates, AT2017fha, SN2017djl, and SN2017djn, have comparable, if not larger, sky support to AT2017gfo ($\iomega\!\approx10^{3}$) but substantially smaller distance overlaps ($\idl\!\approx\!1.6\times 10^{4}$, $1.5\times 10^{4}$, and $1.1\times 10^{4}$), with final scores of
$\iomegadl\approx4.3\times 10^{7}$, $3.3\times 10^{7}$, and $3.2\times 10^{7}$ or factors of $\sim12.1$, $\sim15.5$, and $\sim16.1$ lower, respectively. Thus, this shows that the distance agreement is the dominant contribution for discriminating between individual transients.

Fig.~\ref{fig:margposts} shows the marginalized GW skymap posteriors in right ascension (left), declination (center) and $\dl$ (right), with vertical lines marking the transient candidate positions and the (inferred) distances from the host. Using sky positions alone would leave several candidates tied with ambiguous results. Comparing only the distance information, there are only two candidates that overlap visibly in distance. However, it is after evaluating the full three-dimensional overlap $\iomegadl$ that a single viable counterpart is identified.

As shown in Table~\ref{tab:gw170817stat}, hosts with photometric redshift values centered off of the main posterior will still return nonzero $\idl$ while spectroscopic redshifts away from the GW distance posterior typically give $\idl \approx 0$. This is expected from our mixture model with outlier terms as the broad photometric redshift posteriors overlap the GW line-of-sight values.

For GW170817, the unrelated-source background contains $N_{\rm bg}=32$ pre-merger sources. None of the background sources has a ranking statistic greater than or equal to that of AT2017gfo. Thus, $k=0$ and $p_{\rm cand}=p_{\rm min} = 1/33 =0.030$. Because AT2017gfo is the only post-merger candidate in the search, we do not apply a multiple-candidate correction and $p_{\rm event}=p_{\rm cand}=0.030$. We show $p_{\rm cand}$ as a function $\iomegadl$ on the left side of Fig.~\ref{fig:pvalues}.

The cross-pair calculation provides an additional comparison of the separation of AT2017gfo from the background. Two of the pre-merger sources, AT2016hst and AT2016hrh, were not identified in any of the host galaxy catalogs. Thus, they are excluded from the cross-pair calculation which uses the remaining $N_{\times ,\rm bg} = 30$ sources. We combine the EM and host-galaxy properties of these sources with the positions of the other background sources to obtain $N_{\times}=870$ ordered cross-pairs. 
The largest ranking statistic among the directly observed background sources is $\iomegadl=4.43\times 10^7$ for AT2017fha, as shown in Table~\ref{tab:gw170817stat}. The cross-pair combinations extend to substantially larger values, with a maximum of $3.38\times10^8$, but none of the combinations reaches the ranking statistic value of $\iomegadl\simeq5.19\times10^{8}$ for AT2017gfo, e.g. the largest cross-pair value is a factor of $\simeq1.54$ smaller. Thus, even when the properties and positions from the observed background are recombined, none produces a ranking statistic as large as that of AT2017gfo. The cross-pair calculation is used only as an additional comparison and do not alter $p_{\rm cand}$ or reduce the candidate-level p-value below $p_{\rm min}=0.030$.

\subsection{GW190425}
\label{sec:gw190425}

We apply the same procedure to the candidate transients reported to the TNS for GW190425\footnote{\url{https://www.wis-tns.org/ligo/o1-3/LIGO_-_GW20190425_081805}} within the 99\% probability contour of the GWTC-3 skymap\footnote{\url{https://zenodo.org/records/6513631}} for all sources discovered after the GW trigger time. 
This returns 433 candidates discovered during the post-merger search window spanning $\Delta t_{\rm window}\approx14$ days and 5,101 optical sources discovered during the year prior to the GW trigger time. The pre-merger sources are used for estimating the empirical background population for each candidate's p-value.

The GW190425 localization spans $\sim 8284~\mathrm{deg}^2$ with a distance posterior extending to $\sim 300~\mathrm{Mpc}$, compared with $\sim 28~\mathrm{deg}^2$ at $\sim 40~\mathrm{Mpc}$ for GW170817, and effectively is a searched volume larger by roughly four orders of magnitude. The catalog queries return a larger number of candidate hosts for each transient and at distances where the bulk of the GW distance support lies beyond the GLADE+ completeness horizon. Thus, catalog completeness can be a limiting factor to the recovery of the true host. To address this, we evaluate the missing-host probability $P_{\rm faint}(\hat\Omega)$ (Eq.~\eqref{eq:Pfaint_marginal}) using the local limiting magnitude from each catalog. For GLADE+ we use the published distance-dependent completeness curve, while for PS1-DR2 and LS-DR10, we use the local limiting magnitude or survey-depth estimate entering $C_{\rm eff}(z,\hat\Omega)$ in Eq.~\eqref{eq:completeness_Cz}. After de-duplication across catalogs, the combined queries contribute 23,475 candidate host rows across the set of TNS transients.

The resulting ranking differs qualitatively from the GW170817 case. There is no single GW190425 candidate that clearly outranks the rest of the candidates, as shown in Fig.~\ref{fig:gw190425_margposts}. Instead, the highest-ranked candidates have comparable values of $\mathcal{B}_{\chyp/\shyp\shyp} = \iomegadl$, and several have missing host associations, large $p_\varnothing$, missing host redshifts, or distance posteriors that are only weakly informative. In this instance, the method returns a prioritized list of candidates rather than a confident association, and is the expected outcome for this event given that there is no established counterpart. However, individual candidates could be further distinguished by incorporating additional astrophysical information or priors.

\begin{figure*}
    \centering
    \includegraphics[width=\linewidth]{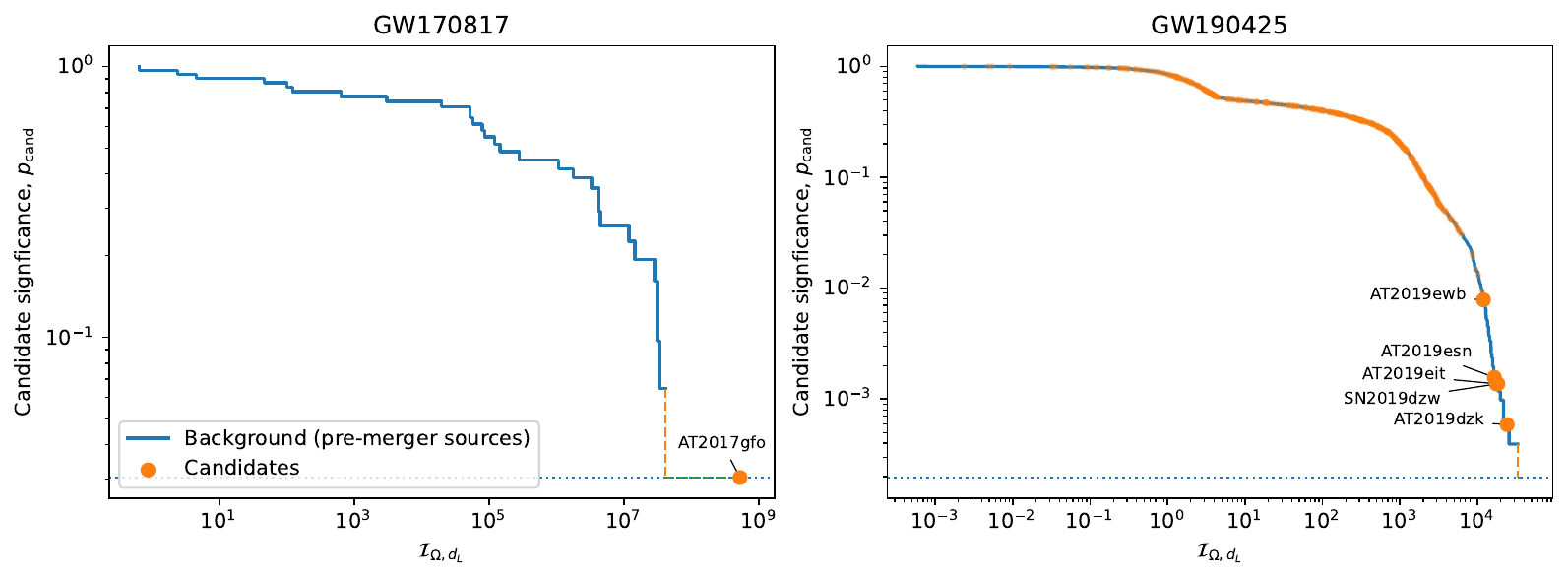}
    \caption{Candidate association significance for GW170817 (left) and GW190425 (right). The figures show the empirical p-value $p_{\rm cand}$ as a function of the ranking statistic $\iomegadl$ using Eq.~\eqref{eq:pvalue_bg} and using the pre-merger backgrounds for each event. The horizontal blue dotted line is the minimum p-value $p_{\rm min} = 1/(N_{\rm bg}+1)$ and the orange scatter points are the post-merger candidates. The dashed line extending from the largest observed background ranking statistic shows the case for $k=0$, where $p_{\rm cand}$ reaches the value of $p_{\rm min}$. For GW170817, AT2017gfo exceeds all of the background sources with $p_{\rm cand}=0.030$. For GW190425, the top 5 candidates for GW190425 are annotated. AT2019dzk has the smallest p-value, $p_{\rm cand}=5.88\times10^{-4}$. The corrected (Holm) p-values are calculated separately and are shown in Table~\ref{tab:gw190425}.}
    \label{fig:pvalues}
\end{figure*}

The right-hand side of Figure~\ref{fig:pvalues} shows the candidate p-value as a function of $\iomegadl$ obtained from the pre-merger background. AT2019dzk has the smallest p-value, $p_{\rm cand}=5.88\times10^{-4}$, with only two background sources having equal or larger ranking statistics.
After applying the Holm correction to all of the 433 candidates, AT2019dzk has the smallest corrected p-value of $p_{\rm corr, (1)}=0.255$. We can then find the event p-value $p_{\rm event}\equiv p_{\rm corr, (1)} =0.255$. Thus, although AT2019dzk has a small p-value $p_{\rm cand}$, accounting for the 433 candidates increases the event-level value and substantially weakens the value reported by $p_{\rm cand}$.

\begin{table*}[t]
\centering
\caption{Summary of the association-significance results for GW170817 and GW190425. $N_{\rm bg}$ is the number of pre-merger sources used to construct the unrelated-source background, $N_{\rm cand}$ is the number of post-merger
candidates examined, and $k$ is the number of background sources with $\iomegadl$ greater than or equal to that of the candidate with the smallest p-value. $p_{\rm min}=1/(N_{\rm bg}+1)$ is the minimum p-value given by the background, $p_{\rm cand}$ is calculated using Eq.~\eqref{eq:pvalue_bg}, and $p_{\rm event}$ is the Holm adjusted p-value defined in Eq.~\eqref{eq:pvalue_event}.}
\label{tab:pvalues}
\begin{tabular}{l@{\quad}c@{\quad}c@{\quad}c@{\quad}c@{\quad}c@{\quad}c@{\quad}c}
\toprule
Event & $N_{\rm bg}$ & $N_{\rm cand}$ & Top Candidate & $k$ & $p_{\rm min}$ & $p_{\rm cand}$ & $p_{\rm event}$ \\
\midrule
GW170817 & 32 & 1 & AT2017gfo & 0 & 0.030 & 0.030 & 0.030 \\
GW190425 & 5101 & 433 & AT2019dzk & 2 & $1.96\times 10^{-4}$ & $5.88\times 10^{-4}$ & 0.255 \\
\end{tabular}
\end{table*}

\subsection{Comparison with existing ranking approaches}
\label{sec:comparison_existing}

We compare the results of our ranking method against existing methods. Table~\ref{tab:comparison_ranks} compares the transient rankings from this work with searched probabilities from \texttt{ligo.skymap} and a galaxy-targeting score computed following \citet{Arcavi2017}. These quantities are not identical estimators, but instead illustrate how they rank the transients relative to $\iomegadl$. The \texttt{ligo.skymap} searched probabilities measure how much GW probability must be searched before reaching a given sky position, distance, or three-dimensional location. The galaxy-targeting score ranks candidate hosts using galaxy properties and the GW localization. The galaxy-targeted strategies used for Swift XRT and UVOT follow-up \citep{Evans2016Follow, Evans2016Optim, EylesFerris2025} use related information for a different stage of the follow-up problem. The methods use the GW localization and galaxy properties to prioritize galaxies or observing fields before an EM counterpart has been identified. The framework presented here is complementary to these strategies, but operate later in the process when transient candidates have already been reported.

For GW170817, the methods that use distance information consistently identify AT2017gfo as the top-ranked candidate. It ranks first by $\iomegadl$, first by $\idl$, first by $P_{\rm vol}$, and first by $S_{\rm tot}$. In contrast, ranking using only sky position does not uniquely recover AT2017gfo. It is ranked ninth by both $\iomega$ and $P_{\rm prob}$. Several sources lie in higher probability regions of the GW skymap, but the host-inferred distances are less consistent with the GW conditional distance posterior. Thus, the recovery of AT2017gfo is driven primarily by the agreement between the GW distance posterior and the host distance, rather than by sky position alone.

The GW170817 comparison also shows the differences between the methods. The three-dimensional searched probability $P_{\rm vol}$ correctly ranks AT2017gfo first, but it loses discrimination among the remaining GW170817 candidates, which are all assigned $P_{\rm vol}\simeq 1$, and is not able to distinguish the candidates. This occurs because $P_{\rm vol}$ is cumulative. The $\iomegadl$ overlap statistic instead preserves the relative differences among candidates, even if they are strongly disfavored by the alternative methods.

For GW190425, the methods do not converge on a single preferred candidate. AT2019dzk ranks first by $\iomegadl$ and second by $P_{\rm vol}$, but ranks fifth by $S_{\rm tot}$ and seventieth by $P_{\rm prob}$. AT2019esn and AT2019ewb rank first and second by $S_{\rm tot}$, but fourth and fifth by $\iomegadl$. The $P_{\rm dist}$ ranking does not identify the same candidates as $\idl$. This is expected because these quantities measure different aspects of the association. In all cases, the quantities compared here should be interpreted as prioritization statistics, not as calibrated association probabilities.

\begin{table*}[t]
\centering
\caption{
Comparison of candidate rankings for GW170817 and GW190425. Candidates are ordered within each event by the host-marginalized $\iomegadl$ calculated in this work. The columns include the values obtained for $\iomega$, $\idl$, and $\iomegadl$ from this work, with the adjacent columns giving the relative rank among all reported candidates for the GW event. The \texttt{crossmatch} columns are the searched probabilities using \texttt{ligo.skymap}: $P_{\rm prob}\equiv\texttt{searched\_prob}$, $P_{\rm dist}\equiv\texttt{searched\_prob\_dist}$, and $P_{\rm vol}\equiv\texttt{searched\_prob\_vol}$, evaluated using the single most probable host galaxy for each transient. $P_{\rm prob}$ is the two-dimensional searched probability, $P_{\rm dist}$ is the sky-marginalized distance CDF evaluated at the host distance, and $P_{\rm vol}$ is the searched probability in the three-dimensional localization volume. $S_{\rm tot}$ is the score found using the galaxy-targeting score computed following \citep{Arcavi2017}. For the overlap statistics and $S_{\rm tot}$, larger values are ranked higher. For $P_{\rm prob}$ and $P_{\rm vol}$, smaller values are ranked higher. The $P_{\rm dist}$ rank is based on closeness to 0.5, corresponding to consistency with the median of the sky position marginalized GW distance distribution.
}
\label{tab:comparison_ranks}
\begin{adjustbox}{max width=\textwidth, center}
\begin{tabular}{@{}ll@{\quad}cc@{\quad}cc@{\quad}cc@{\quad}cc @{\quad}cc @{\quad}cc ccc@{}}
\toprule
Event & Candidate
& \multicolumn{6}{c}{This work}
& \multicolumn{6}{c}{\texttt{ligo.skymap crossmatch}}
& & \multicolumn{2}{c}{Galaxy targeting} \\
\cmidrule(lr){3-8}
\cmidrule(lr){9-14}
\cmidrule(lr){16-17}
& &
$\iomega$ & Rank$_\Omega$ &
$\idl$ & Rank$_{D_L}$ &
$\iomegadl$ & Rank$_{\Omega,D_L}$ &
$P_{\rm prob}$ & Rank$_{\rm prob}$ &
$P_{\rm dist}$ & Rank$_{\rm dist}$ &
$P_{\rm vol}$ & Rank$_{\rm vol}$ & &
$S_{\rm tot}$ & Rank$_S$ \\
\midrule
GW170817 & AT2017gfo & 1375 & 9 & $3.77 \times 10^5$ & 1 & $5.19\times 10^8$ & 1 & 0.556 & 9 & 0.495 & 1 & 0.39 & 1 & & 1.00 & 1 \\
GW170817 & AT2017fha & 2620 & 4 & $1.69\times 10^4$ & 3 & $4.43\times 10^7$ & 2 & 0.070 & 2 & 1.00 & 5 & 1.00 & 5 & & $\ll 1$ & 6 \\ 
GW170817 & SN2017djn & 3011 & 1 & $1.14\times 10^4$ & 10 & $3.45\times 10^7$ & 3 & 0.002 & 1 & 1.00 & 14 & 1.00 & 5 & & $\ll 1$ & 15 \\
GW170817 & SN2017djl & 2207 & 7 & $1.51\times 10^4$ & 4 & $3.33\times 10^7$ & 4 & 0.317 & 7 & 1.00 & 6 & 1.00 & 5 & & $\ll 1$ & 7 \\
GW170817 & SN2017ddy & 2680 & 2 & $1.16\times 10^4$ & 7 & $3.10\times 10^7$ & 5 & 0.112 & 3 & 1.00 & 13 & 1.00 & 5 & & $\ll 1$ & 13 \\
\midrule
GW190425 & AT2019dzk & 3.46 & 141 & $7.04\times 10^3$ & 5 & $2.43 \times 10^4$& 1 & 0.40 & 70 & 0.12 & 12 & 0.17 & 2 & & 0.09 & 5\\
GW190425 & SN2019dzw & 3.35 & 148 & $5.47\times 10^3$ & 8 & $1.84 \times 10^4$ & 2 & 0.80 & 161 & 0.26 & 16 & 0.57 & 9 & & 0.04 & 7 \\
GW190425 & AT2019eit & 2.32 & 220 & $7.58\times 10^3$ & 4 & $1.76 \times 10^4$ & 3 & 0.77 & 154 & 0.44 & 19 & 0.59 & 11 & & 0.10 & 4\\
GW190425 & AT2019esn & 3.55 & 132 & $4.67\times 10^3$ & 10 & $1.66 \times 10^4$ & 4 & 0.56 & 103 & 0.18 & 13 & 0.53 & 7 & & 0.18 & 1 \\
GW190425 & AT2019ewb & 3.03 & 177 & $3.96\times 10^3$ & 11 & $1.20 \times 10^4$ & 5 & 0.60 & 109 & 0.67 & 28 & 0.50 & 6 & & 0.13 & 2\\
\bottomrule
\end{tabular}
\end{adjustbox}
\end{table*}

\vspace{5mm}

\subsection{Considerations and caveats}
\label{sec:catalog_caveats}

\begin{figure}[t!]
    \centering
    \includegraphics[width=\linewidth]{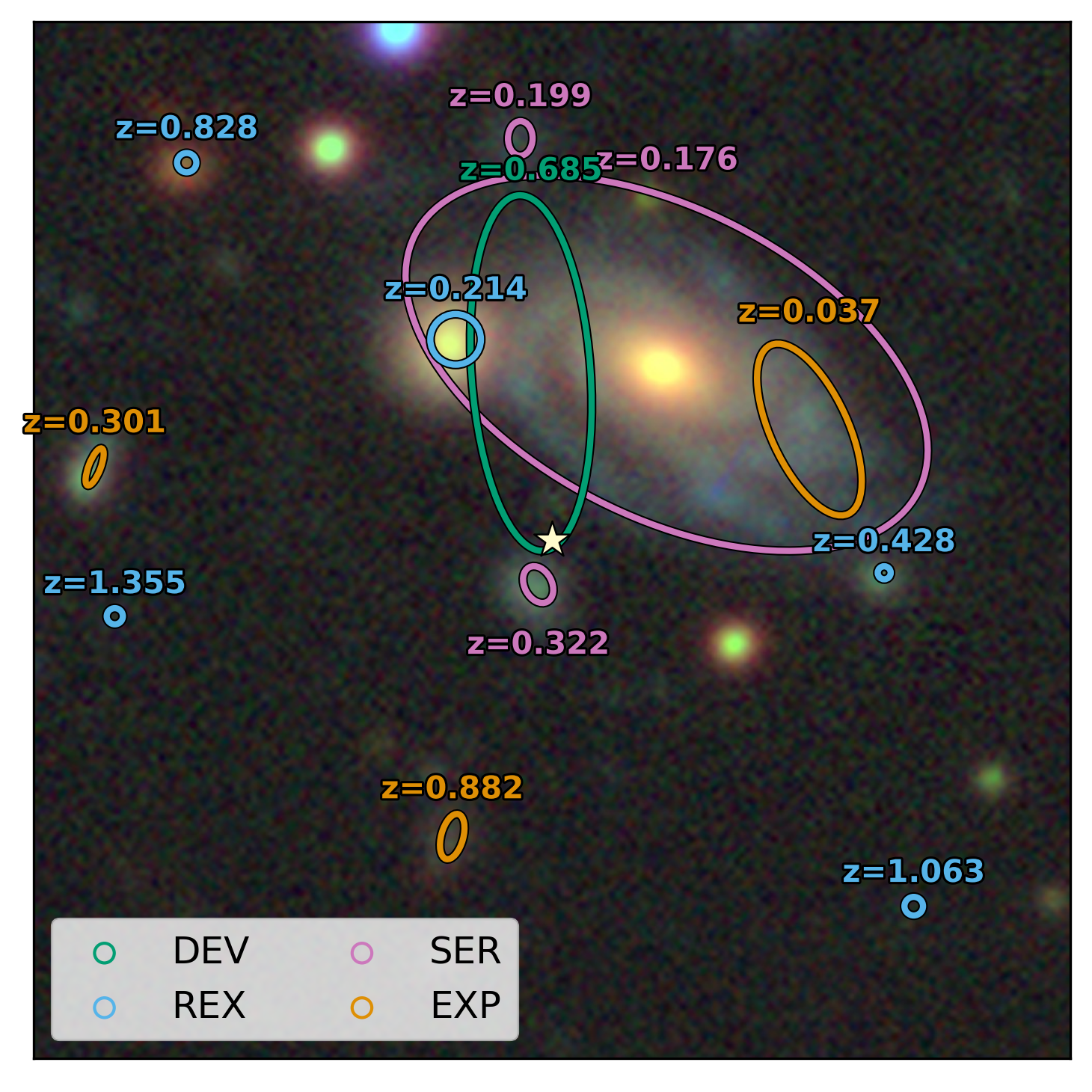}
    \caption{DESI Legacy Survey DR10 cutout centered on a transient candidate (yellow star). The ellipses are cataloged galaxy model fits: DEV (de Vaucouleurs, green), REX (round exponential, blue), SER (S\'ersic, magenta), and EXP (exponential, orange). Text labels give the cataloged photometric redshift for each entry. The bright central galaxy is represented by multiple deblended model components with inconsistent $\zphot$ values ($z=0.037$, $0.176$, $0.322$, $0.685$, $0.214$), which illustrates how wide-field catalogs can return several overlapping candidates around a single host. Naively choosing the ellipse containing the transient (green, $z=0.685$) would misassociate the transient.}
    \label{fig:galconfusion}
\end{figure}

Within our chosen catalogs the low-redshift $\zphot$ estimates can be highly unreliable or entirely incorrect. In LS-DR10 and PS1-DR2 we often find multiple detections around galaxies, sometimes dozens, where each detection reports its own $\zphot$. In some instances, the central galaxy in LS-DR10 may be tagged as a Gaia duplicated source. When this occurs, photometry may not be run on the source and may be entirely omitted from the $\zphot$ analysis, even though nearby deblended detections remain. In one representative field, shown in Fig.~\ref{fig:galconfusion}, there are four sources with distinct photometric redshift associated to the same host, with values ranging from $z=0.037$ to $z=0.685$. In this instance, a na\"ive choice may lead to the green ellipse being marked as the true host. This is due to the source identified in the catalog as a DEV-type galaxy with $z=0.685$ that encloses the location of the transient. However, this source appears to be a false detection which would lead to a biased and incorrect redshift inference, and the transient being downweighted to a more distant galaxy outside of the GW distance range.

This motivates incorporating a crossmatching stage of the LS-DR10 and PS1-DR2 hosts against catalogs with bright, nearby host galaxies from NED, NED-LVS, GLADE+, or REGALADE \citep{Tranin2026}. In practice, GLADE+ and nearby catalogs (GWGC, PGC/HyperLEDA) perform better for $z\!\lesssim\!0.03$ because of targeted completeness in the local volume. 

\section{Discussion}
\label{sec:discussion}

Our GW170817 case study shows that the ranking statistic in Eq.~\eqref{eq:jointoverlap_delta} is effective for GW-EM association. The sky overlap term $\iomega$ identifies transients consistent with the GW skymap at sub-arcsecond precision, while the distance term $\idl$ quantifies agreement between the GW line-of-sight distance posterior and the EM distance inferred from the host redshift. In practice, $\idl$ is the dominant discriminator that sets the overall ranking, and without this term the ranking is ambiguous, as shown in Table~\ref{tab:gw170817stat} and in the conditional posteriors of Figs.~\ref{fig:margposts} and \ref{fig:dls_gw170817}. 

Because Eq.~\eqref{eq:distoverlap} contains the inverse prior $\pi(\dl)$, the value of $\idl$ scales with the prior's normalization, which is ultimately scaled by the integrand limits and the prior range. Therefore, to compare transient candidates against each other for a single GW event and to make relative comparisons between GW events, we fix a single, shared luminosity distance prior. While we choose to adopt a uniform in comoving volume prior on $[10,10^4]$\,Mpc throughout Sec.~\ref{sec:distanceoverlap}, any reasonable range can be substituted in as long as it remains consistent across event rankings and that the public skymaps are reweighted to the same $\pi(\dl)$.

The DLR-normalized offset $\rdlr$ provides a morphology-aware alternative to the angular separation $\Delta\theta$, as the latter are not directly comparable across hosts with different sizes, ellipticities, and orientations. For example, an angular separation can correspond to a location well outside the light profile of a compact dwarf galaxy, but still lie within the light profile of an extended elliptical galaxy. Normalizing angular separation by DLR rescales the separation as directionally-dependent offsets that are directly comparable across galaxies with different angular sizes, ellipticities, and orientations. 

The same normalization also determines how astrometric uncertainty enters the offset likelihood. When combined with the Rice astrometric kernel, the uncertainties propagate approximately as $\sigma_{\rdlr}\simeq\sigma_\theta/\mathrm{DLR}$ which produces a narrower offset likelihood for well-resolved hosts and broader for compact or poorly measured hosts. This helps to down-weight chance alignments at large $\rdlr$ while preserving genuine small-offset associations. We show this with candidate AT2017bej in Fig.~\ref{fig:hostassoc_dlr}, where we see that in fields with multiple plausible hosts, the morphology-aware likelihood prevents large angular-separation associations from being favored solely because the galaxy is extended, while preserving small-$\rdlr$ associations as plausible hosts.

Photometric redshifts at low-$z$ are often biased, duplicated, or non-Gaussian in wide-field catalogs. We therefore search for available $\zspec$ values via NED/NED-LVS and replace them if available. Otherwise, we adopt a calibrated mixture model that is mapped to $\dl$ under our fiducial cosmology. This suppresses spurious peaks in $\idl$ when long photometric redshift tails have probability that crosses over the GW posterior and provides conservative support where $\zphot$ is uncertain. Marginalizing peculiar velocity with an assumed $\sigma_v=250~\kms$ further broadens EM distance posteriors at $z\lesssim0.1$ to account for uncertainties in the local flow \citep{Blake2025}.

We also include the propagation of measurement uncertainties in astrometry, morphology, and redshift via Monte Carlo sampling. This is to avoid artificially narrow $\iomega$ or $\idl$ values from point-estimate evaluations. While Sec.~\ref{sec:missinghosts} includes an explicit first-order catalog-completeness term, the result still depends on the adopted completeness model, the assumed Schechter luminosity function, and the local limiting-magnitude estimates. Other systematic uncertainties include Monte Carlo sampling noise and biases in the choice of peculiar velocity and offset priors.

Galaxy catalog incompleteness can still introduce systematic bias because hosts absent from the catalogs cannot be assigned directly to a transient. They may be absent because of intrinsic properties, such as ultra-faint dwarf galaxies or cluster members below survey detection thresholds, or because of observational effects such as Galactic extinction, surface-brightness limits, masking, or crowding. The missing host term reduces over-confident host assignments in these regimes, but it does not remove the need for deeper follow-up imaging when the true host is absent from the available catalogs.

When the ranking does not yield a single dominant candidate, as for GW190425 where the top-ranked candidate is only marginally larger than the remaining candidates, the statistic should be interpreted as a prioritization metric rather than as a detection claim. Candidates whose $\iomegadl$ values are consistent within their Monte Carlo uncertainties, or differ by less than a factor of a few, should be treated as ambiguous and should require additional photometric, spectroscopic, or temporal information. In such cases, the method identifies which candidates are most consistent with the GW localization and host catalog information, but it does not by itself establish a secure counterpart.

The framework can be extended by integrating additional astrophysical host priors. For example, cataloged galaxies can be weighted by stellar mass, star-formation rate, or luminosity when these properties are reliably measured \citep{Palfi2025, Artale2020}. They could be folded directly into the host galaxy prior in Eq.~\eqref{eq:hostmatch} without modifying the core ranking statistic.

A second extension involves the transient itself. Multi-epoch photometry, the rise rate, peak brightness, color, and decay timescale can be compared to kilonova or GRB afterglow templates \citep[e.g.][]{Coughlin2018, Setzer2019}. Even sparse photometry can provide some discrimination. For example, a slowly-evolving transient at $t_{\rm GW} + 5$\, days is unlikely to be a kilonova, while a fast-fading red source at $t_{\rm GW} + 5$\, days is a stronger candidate. These can enter as an additional likelihood term which is conditional on a model and viewing angle \citep[e.g.][]{Chen2019}. However, care must be taken as such a term is highly dependent on source model assumptions and given the degeneracy of the shared parameters, i.e., $\dl$--inclination and $\dl$--mass degeneracy. This is particularly the case if one were to rely on the low-latency parameter estimates.

In this work, we estimate the statistical significance of the ranking statistic using a background distribution of transients observed before the GW trigger time. This returns a p-value for each candidate. For events with more than one candidate, the p-value is corrected for the total number of candidates considered in the search. These p-values quantify how extreme the ranking statistic is compared to the background. They are conditional on the background population and candidate selection and should not be interpreted as posterior probabilities of association or as joint GW-EM false-alarm rates.

A full posterior odds calculation would require the prior odds ratio between the common-source and unrelated source hypotheses, see Eq.~\eqref{eq:postodds}. Similarly, converting the significance to a joint false alarm rate requires information about the rate of unrelated optical transients (contaminants), duration and selection function of the survey, and rate of image-level false positives, in addition to the astrophysical GW event rate, and the GW false alarm rate. These quantities are not calibrated here. The empirical p-value developed in Sec.~\ref{sec:pvalue} provides a significance measure for the ranking statistic without requiring these additional rate assumptions. A full posterior association probability or joint false-alarm rate is left to future work.

Additionally, our choice of priors may have a noticeable effect on our final results. Allowing $\sigma_v$ to vary by host type would widen $p(\dl\midbar\dem)$ for those hosts. However, this would not change the qualitative result that $\idl$ dominates over $\iomega$. We also expect that differences in morphology across catalogs can introduce small shifts in $\mathrm{DLR}$. By working in the normalized coordinates of $\rdlr$, this helps to mitigate this systematic, but it does not eliminate it.

\subsection{Applications to low-latency infrastructure}
We also remark that this approach is suitable for a low-latency implementation. For a fixed GW skymap, all of the candidates can be independently evaluated and can be parallelized. Once the relevant host catalog information has been retrieved, the computational cost scales approximately with the number of transient candidates, the number of candidate host galaxies per transient, and the number of Monte Carlo uncertainty realizations. In practice, the total computational wall time is not determined by the GW sky area alone. The dominant contribution to the latency is in querying the remote galaxy catalogs, which depend on the external service and network conditions. Given the large search cone around each transient and along the GW distance line of sight, a single unoptimized query can return tens to hundreds of thousands of sources. The total number can be reduced by submitting server-side filtering, see Appendix~\ref{app:recipes}, but this can be a limitation if the query service does not support this feature.

Hosting a galaxy catalog locally can substantially reduce the latency associated with remote catalog queries. Several catalogs are already well suited to nearby-GW follow-up, including GLADE+ \citep{Dalya2022}, HECATE \citep{Kovlakas2021}, NED-LVS \citep{Cook2023}, REGALADE \citep{Tranin2026}, and upcoming UpGLADE \citep{DalyaUpGLADE, Brozzetti2024}. However, these resources are primarily optimized for the nearby Universe and GW-targeted applications. In practice, many transients reported to TNS are selected because they fall within the 90\% contour of the two-dimensional GW localization, irrespective of their distance. For many of the transients, they may require deeper, more general-purpose galaxy catalogs.

The need for rapid candidate prioritization is becoming more critical. This is especially the case for new facilities, i.e., optical survey instruments that are significantly deeper, like Rubin \citep{Ivezic2019}, upcoming \textit{extremely} wide-field facilities, such as Argus Array \citep{Law2022} and Digital Telescope \citep{Pollaco2026}, and next-generation GW facilities, such as Cosmic Explorer \citep{Reitze2019} or Einstein Telescope \citep{Punturo2010}. It is already impractical to exhaustively characterize every spatially-coincident transient. The limited fraction of classified follow-up transients further motivates a ranking method that does not depend on transient classification. With GW and EM facility upgrades, the increase in GW detection rate and accessible volume together with the order-of-magnitude increase in optical transient detections makes exhaustive characterization increasingly impractical. Thus, a ranking and significance framework based on information that is immediately available after a GW alert can provide an initial filter for allocating the limited follow-up resources. The method presented here is particularly suited for this application as it does not require any transient classification or lightcurve information, and the rankings can be updated in real time as new transients are reported.

\vspace{1em}

\section{Conclusions}
\label{sec:conclusion}

We presented a Bayesian ranking framework for associating EM transients with GW events that combines (i) the three-dimensional GW skymap, (ii) a morphology-aware host association in normalized DLR units, and (iii) host-inferred distance posteriors with peculiar velocity effects considered. The final combined statistic, $\mathcal{B}_{\chyp/\shyp\shyp} \simeq \iomegadl\simeq \iomega\,\idl$, provides an interpretable ranking statistic for associating and ranking EM transient candidates with GW events. In addition, we estimate a significance for each candidate using the unrelated-source background and account for multiple post-merger candidates using the Holm correction.

When applied to GW170817, the method identifies AT2017gfo as the top-ranked candidate by a factor of $\sim 12$ and correctly recovers NGC\,4993 as the host. AT2017gfo exceeds all of the background sources and $p_{\rm cand}=p_{\rm event}=0.030$. For GW190425, $p_{\rm cand}=5.88\times10^{-4}$. After accounting for all 433 candidates, this is corrected to $p_{\rm corr}=0.255$. Because this is the smallest corrected p-value, we find $p_{\rm event}=0.255$. The method finds that none of the candidates provide statistically significant evidence for to be associated to GW190425.

Additionally, this approach is directly applicable to ongoing and archival searches \citep{Ackley2026b}. It can operate on TNS reported candidates or on Rubin alert streams distributed by community brokers, such as ALeRCE \citep{Forster2021}, ANTARES \citep{Matheson2021}, Lasair \citep{Williams2024}, or FINK \citep{Moller2021}, among others, or alongside contextual classifiers like \textit{Sherlock} \citep{Sherlock}, \texttt{moriarty} \citep{Killestein2023}, or \texttt{ORACLE} \citep{Oracle2025}. In particular, this type of automated candidate prioritization will become increasingly important for next-generation GW facilities, where the expected increase in the number of GW alerts and unrelated optical transient candidates can make exhaustive follow-up practically impossible.

It is also adaptable to a broad array of transients requiring host association, such as nearby GRBs with large localization uncertainty regions from Fermi-GBM \citep{Meegan2009, Connaughton2015}, where existing low-redshift galaxy catalogs provide reasonable completeness. This framework was recently applied to assessing the host association probability for GRB\,250818B \citep{Belkin2026}. For higher-redshift events, the galaxy catalog completeness becomes the leading limitation and dedicated deeper follow-up imaging is required to find an underlying host.

Future extensions of this work include calibrating offset prior hyperparameters with a larger sample; transient photometric evolution; weighting host galaxies by properties such as luminosity, stellar mass, or star-formation rate; and the inclusion of prior odds to obtain a full odds ratio. These additions should further improve the efficiency of EM follow-up, reduce the number of false positives that need to be verified, and increase the likelihood of discovering additional EM counterparts, particularly those associated with sub-threshold GW events.

\section*{Acknowledgements}
I am deeply grateful to the anonymous referees for their exceptionally thoughtful, constructive, and insightful review. Their detailed comments and suggestions substantially strengthened the statistical methodology and presentation, and significantly improved both the scope and clarify of the manuscript. I would like to thank Greg Ashton and Naresh Adhikari for discussions regarding the statistical development of this work, Federico Stachurski for discussions on cosmological inference, Kenneth Duncan for discussions on the Legacy Survey catalog, and Danny Steeghs for helpful feedback on the manuscript. This work was supported by the The Science and Technology Facilities Council (STFC) via grant number ST/X001121/1.

This research has used data or software obtained from the Gravitational Wave Open Science Center (gw-openscience.org), a service of LIGO Laboratory, the LIGO Scientific Collaboration, the Virgo Collaboration, and KAGRA. LIGO Laboratory and Advanced LIGO are funded by the United States National Science Foundation (NSF) as well as the Science and Technology Facilities Council (STFC) of the United Kingdom, the Max-Planck-Society (MPS), and the State of Niedersachsen/Germany for support of the construction of Advanced LIGO and construction and operation of the GEO600 detector. Additional support for Advanced LIGO was provided by the Australian Research Council. Virgo is funded, through the European Gravitational Observatory (EGO), by the French Centre National de Recherche Scientifique (CNRS), the Italian Istituto Nazionale di Fisica Nucleare (INFN), and the Dutch Nikhef, with contributions by institutions from Belgium, Germany, Greece, Hungary, Ireland, Japan, Monaco, Poland, Portugal, and Spain. The construction and operation of KAGRA are funded by Ministry of Education, Culture, Sports, Science and Technology (MEXT), Japan Society for the Promotion of Science (JSPS), National Research Foundation (NRF), and Ministry of Science and ICT (MSIT) in Korea, Academia Sinica(AS) and the Ministry of Science and Technology (MoST) in Taiwan.

This research uses services or data provided by the Astro Data Lab, which is part of the Community Science and Data Center (CSDC) Program of NSF NOIRLab. NOIRLab is operated by the Association of Universities for Research in Astronomy (AURA), Inc. under a cooperative agreement with the U.S. National Science Foundation.

The Photometric Redshifts for the Legacy Surveys (PRLS) catalog used in this paper was produced thanks to funding from the U.S. Department of Energy Office of Science, Office of High Energy Physics via grant DE-SC0007914. 
This research has made use of the NASA/IPAC Extragalactic Database, which is funded by the National Aeronautics and Space Administration and operated by the California Institute of Technology.
This publication makes use of data products from the Wide-field Infrared Survey Explorer, which is a joint project of the University of California, Los Angeles, and the Jet Propulsion Laboratory/California Institute of Technology, funded by the National Aeronautics and Space Administration.
This research has made use of data obtained from the SuperCOSMOS Science Archive, prepared and hosted by the Wide Field Astronomy Unit, Institute for Astronomy, University of Edinburgh, which is funded by the UK Science and Technology Facilities Council.

\appendix
\setcounter{table}{0}
\renewcommand{\thetable}{A\arabic{table}}
\renewcommand{\theHtable}{A\arabic{table}}
\onecolumngrid

\section{Catalog query and filtering recipes}
\label{app:recipes}

We list the filters and joins used to build the host-galaxy sample from
LS-DR10, PS1-DR2, PS1-STRM/WISE-PS1-STRM, and GLADE+. Each catalog row includes the chosen \texttt{best\_band}, \texttt{best\_band\_snr}. The corresponding \texttt{best\_*} photometry and shape fields are used downstream for evaluating $\rdlr$ Eq.~\eqref{eq:rdlr} and the host likelihood Eq.~\eqref{eq:spatial_off_gal}. We use AT2017gfo as the reference transient where we perform a radial query centered on the transient within $1040\arcsec = 0.288889^\circ$.
\subsection{DESI Legacy Survey DR10 (LS-DR10)}
Queries to the DESI Legacy Survey DR10 catalog are made through the Astro Data Lab Python client \texttt{datalab}\footnote{\url{https://github.com/astro-datalab/datalab}}
\begin{itemize}
\item \textbf{Source-type and data quality filter}: We exclude PSF sources and any objects carrying specific LS-DR10 \texttt{fitbits} quality flags\footnote{ \url{https://www.legacysurvey.org/dr10/bitmasks/}}$^{,}$\footnote{ \url{https://github.com/legacysurvey/legacypipe/blob/DR10.0.12/py/legacypipe/bits.py}}:
\begin{verbatim}
    MASK = (FORCED_POINTSOURCE | HIT_RADIUS_LIMIT | HIT_SERSIC_LIMIT |
            FROZEN | RUNNER | GAIA_POINTSOURCE | ITERATIVE)
    keep = (type != 'PSF') and ((fitbits & MASK) == 0)
\end{verbatim}
where we only check against the following flags listed in Table~\ref{tab:lsflags}.

\begin{table}[ht]
\centering
\caption{LS--DR10 \texttt{FITBITS} excluded by our mask. We retain sources with
\(\texttt{type} \neq \texttt{'PSF'}\) and with none of the listed bits set}
\setlength{\tabcolsep}{10pt}
\begin{tabular}{lll}
\hline
Flag & Hex & Description \\
\hline
\texttt{FORCED\_POINTSOURCE} & 0x0001 & Source was forced to be PSF-like. \\
\texttt{HIT\_RADIUS\_LIMIT} & 0x0004 & Fit hit radius limit. \\
\texttt{HIT\_SERSIC\_LIMIT} & 0x0008 & S\'ersic constraints reached. \\
\texttt{FROZEN} & 0x0010 & Parameters frozen (not fit). \\
\texttt{RUNNER} & 0x0800 & Runner source (non-standard fit). \\
\texttt{GAIA\_POINTSOURCE} & 0x1000 & Gaia-matched point source. \\
\texttt{ITERATIVE} & 0x2000 & Iterative fit flag set. \\
\hline
\end{tabular}
\label{tab:lsflags}
\end{table}
\item \textbf{Pick the best band} (among $g,r,z$) as the band with the maximum SNR.
\item \textbf{Require finite values of}:
\begin{quote}
  morphology: \texttt{shape\_r},  \texttt{shape\_e1},  \texttt{shape\_e2} \\
  magnitude: \texttt{mag\_\{best\}} \\
  flux fraction: \texttt{fracflux\_\{best\}}
\end{quote}
\item \textbf{Crossmatch to additional catalogs to verify source type}, including DESI Legacy photometric redshift, GAIA, 2MASS-XSC, 
\begin{verbatim}
    ls_dr10.tractor LEFT JOIN:
        ls_dr10.photo_z
        ls_dr10.x1p5__tractor__nsc_dr2__object
        ls_dr10.x1p5__tractor__gaia_dr3__gaia_source
        nsc_dr2.x1p5__object__twomass__xsc
        gaia_dr3.gaia_source
\end{verbatim}
\end{itemize}
\subsection{Pan-STARRS DR2 (PS1-DR2)}
Queries to the Pan-STARRS1 DR2 are made through a MAST API search query\footnote{\url{https://ps1images.stsci.edu/ps1_dr2_api.html}}. Following \citet{Magnier2020}, \texttt{primaryDetection} and \texttt{bestDetection} should not be used for either DR1 or DR2 queries.
\begin{itemize}
\item \textbf{Astrometry}: Use \texttt{raMean}, \texttt{decMean} (rather than \texttt{raStack}, \texttt{decStack} as the stack positions have unreliable error estimates) and require \texttt{raMeanErr $<$ 1000} and \texttt{decMeanErr $<$ 1000}.
\item \textbf{Quality}: Following \citet{Flewelling2020}, we exclude sources that do not meet the quality flag requirements: 
\begin{verbatim}
    qualityFlag != 128
    (gpsfQfPerfect | rpsfQfPerfect | zpsfQfPerfect) >= 0.95
\end{verbatim}
where \texttt{psfQfPerfect} is the PSF-weighted fraction of pixels totally unmasked for the filter stack detection.
\item \textbf{Stack selection}: do not filter on \texttt{primaryDetection} or \texttt{bestDetection}. Instead, we set a flag on the filter \texttt{infoFlag3} 
\begin{verbatim}
    (ginfoFlag3 | rinfoFlag3 | zinfoFlag3) = STACK_TOTAL
    STACK_TOTAL = 196608 = 65536 (STACK_PRIMARY) + 131072 (STACK_PHOT_SRC)
\end{verbatim}
where \texttt{STACK\_PRIMARY} is set for stack measurements that are the primary measurement and \texttt{STACK\_PHOT\_SRC} is set for the best measurement of an object in a given filter \citep{Flewelling2020}.
\item \textbf{Pick the best band}: choose the band with the maximum SNR.
\item \textbf{Require finite values of}
\begin{quote}
  second moments: \texttt{momentXX, momentXY, momentYY}\\
  non-zero Kron radius: \texttt{\{band\}KronRad}\\
  finite Kron magnitude: \texttt{\{band\}KronMag}
\end{quote}
\end{itemize}

\subsection{GLADE+}
Queries to the GLADE+ catalog are made through the Python package \texttt{astroquery}\footnote{\url{https://github.com/astropy/astroquery}}. Queries to the SuperCOSMOS database are made through \texttt{HTTP/POST} requests\footnote{\url{http://ssa.roe.ac.uk/sql.html}}
\begin{itemize}
\item \textbf{Crossmatch against source catalogs to obtain morphology}. Prioritize in order of PGC, HyperLEDA, 2MASS-XSC, WISExSuperCOSMOS and take the first available.
\item For matched WISExSuperCOSMOS sources, query morphologies via SSA (SuperCOSMOS).
\end{itemize}

\subsection{NED and NED-LVS}
Queries to NED and NED-LVS are made before the final host is chosen. This is so that the photometric and spectroscopic redshift values are checked against potential spectroscopic matches. Queries to the NED catalog are made through the Python package \texttt{astroquery} while queries to the NED-LVS are made through a local instance of the database\footnote{\url{https://ned.ipac.caltech.edu/NED::LVS/}}.
\begin{itemize}
\item \textbf{Crossmatch the GLADE+, LS-DR10, or PS1-DR2 candidate hosts}. Crossmatch sources within 2 arcseconds of the candidate host. If more than one result is returned, prefer the object with an NGC or Messier designation, otherwise choose the source with the small angular separation. We exclude entries that are themselves classified as supernovae (\texttt{Type != SN}) or that correspond to the transient label. 
\item \textbf{Require finite redshift values}. If there is a corresponding velocity reported, $cz$, prefer this value.
\end{itemize}

\subsection{Gaia crossmatch and foreground-star rejection}
\label{app:gaia}

Gaia DR3 is used to identify and remove foreground stellar contaminants from the set of host candidates. We crossmatch each source to Gaia DR3 and retain the best Gaia match. We then apply the foreground-star criteria of \citet{Barmby2023}, using the non-zero parallax and significant proper-motion tests, corresponding to their Eqs.~1 and 2. We do not apply their Eqs.~3 and 4 as those require a known systemic proper motion for the background galaxy, which is not typically available for all catalogs. We use the conservative threshold for the criteria of $n=10$.

In addition, we require a high probability of being a single star and a bright Gaia magnitude before removing the object from the host candidate list. Specifically, we remove a source only if it satisfies the foreground flag, has $P_{\rm SS}\ge0.999$, and has $G\le20.5$. The $G=20.5$ threshold follows the characteristic magnitude at which foreground sources become an important fraction of Gaia detections \citep{Barmby2023}. This combination avoids removing compact background galaxies that have a Gaia match but are not confidently foreground stars.

\twocolumngrid
\bibliographystyle{apsrev4-2}
\bibliography{references}

\end{document}